\documentclass[aps,twocolumn,secnumarabic,nobalancelastpage,amsmath,amssymb,
nofootinbib]{revtex4-1}
\pdfoutput=1 
\usepackage{dcolumn}
\usepackage{bm}
\usepackage[T1]{fontenc}
\usepackage[latin1]{inputenc}
\usepackage[active]{srcltx}
\usepackage{ae,aecompl}
\usepackage[brazil]{babel}
\usepackage{caption}
\usepackage{subcaption}
\usepackage{graphicx}
\usepackage{amsmath}
\usepackage{color}
\usepackage{braket}
\usepackage[labelformat=parens,labelsep=quad,skip=3pt]{caption}

\begin{document}

\title{Universality and thermoelectric transport properties of quantum dot systems}

\author{D. F. Aranguren-Quintero$^1$, E. Ramos $^2$, J. Silva-Valencia $^1$, M. S. Figueira $^3$, L. N. Oliveira$^4$ and  R. Franco$^1$}

\affiliation{$^1$ Departamento de F\'{\i}sica, Universidad Nacional de Colombia (UNAL), A. A. 5997, Bogot\'{a}, Colombia.\\
$^2$ Divisi\'{o}n de Ciencias B\'{a}sicas, Fundaci\'{o}n Universidad de America, Bogot\'{a}, Colombia.\\
$^3$ Instituto de F\'{i}sica-Universidade Federal Fluminense (IF-UFF), Av. Litor\^{a}nea s/n, CEP:24210-346, Niter\'{o}i, Rio de Janeiro, Brazil.\\
$^4$ Instituto de F\'{i}sica de S\~{a}o Carlos-Universidade de S\~{a}o Paulo (IFSC-USP), 369 S\~{a}o Carlos, S\~{a}o Paulo, Brazil.}

\begin{abstract}
We discuss the temperature-dependent thermoelectric transport properties of semiconductor nanostructures comprising a quantum dot coupled to quantum wires: the thermal dependence
of the electrical conductance, thermal conductance, and thermopower. We explore the universality of the thermoelectric properties in the temperature range associated with the 
Kondo crossover. In this thermal range, general arguments indicate that any equilibrium property's temperature dependence should be a universal function of the ratio
 $T^{*}=T/T_{K}$, where $T_{K}$ is the Kondo temperature. 

Considering the particle-hole symmetric, spin-degenerate Anderson model, the zero-bias electrical conductance has
already been shown to map linearly onto a universal conductance through a quantum dot embedded or side-coupled to a quantum wire. Employing  rigorous renormalization-group 
arguments, we calculate universal thermoelectric transport coefficients that allow us to extend this result to the thermopower and the thermal conductance. We present numerical 
renormalization-group results to illustrate the physics in our findings. 

Applying the universal thermoelectric coefficients to recent experimental results of the electrical conductance and thermo-voltages versus $V_{gate}$, at different temperatures 
in the Kondo regime, we calculate all the thermoelectric properties and obtain simple analytical fitting functions that can be used to predict the experimental results of 
these properties. However, we cannot check all of them, due to the lack of available experimental results over a broad temperature range.
\end{abstract}


\date[Date: ]{\today}
\maketitle

\section{Introduction}
\label{sec1}

The discovery of the Seebeck and Peltier effects in different metal junctions at the beginning of the $19^{th}$ century gives rise to a branch of physics cakked ``Thermoelectric" 
(TE) \cite{Rosa16}. Seebeck observed that when two different metals joined together (thermocouple) with the junctions maintained at different temperatures, a voltage difference 
is generated proportional to the temperature variation between the couple's ends. Some times later, Peltier observed that when an electric current flows through the Seebeck 
device,  heat is either absorbed or rejected depending on the direction of the current along the circuit. Today, the Peltier effect is the basis for many TE refrigeration 
devices, and the Seebeck effect is the basis for TE power generation devices \cite{Tritt2002}.

Ioffe's prediction in the fifties that doped semiconductors could exhibit relatively large thermoelectric effects \cite{Joffe_1959} had a strong impact on the area of 
thermoelectric materials. It was followed step  by step with the discovery that a thermo-junction between $p-$type $Bi_{2}Te_{3}$ and bismuth exhibits the maximum temperature 
difference of  $26^\circ$C and $40^\circ$C between ($p-$ and $n-$) types $Bi_{2}Te_{3}$  \cite{Wright1958}. This compound has dominated the whole field of thermoelectric 
materials; more specifically, the alloys of $Bi_{2}Te_{3}$  with $Sb_{2}Te_{3}$ for $p-$type and $Bi_{2}Se_{3}$ for $n-$type compounds have the highest $ZT$ (see Eq. \ref{ZT}), 
at around room temperatures, when compared to any other known material \cite{Witting2019}, and up until now, it is the working material for most Peltier cooling devices and 
Seebeck thermoelectric generators. 

Most state-of-the-art thermoelectric materials have their dimensionless thermoelectric figure of merit, $ZT$, in the interval $ZT\simeq [1-2.5]$ (see Fig. 2 of the reference  
\cite{Jian17}), which is well below the Carnot efficiency \cite{Benenti2017}. However, the advent of nanotechnology opens up new possibilities for increasing $ZT$, mainly due to 
the level quantization and the Coulomb interaction, leading to essential changes in the system's thermoelectric properties. Some promising compounds are the topological insulators 
(TIs) and Weyl and Dirac's semi-metals, characterized by nontrivial topological orders. The new characteristic of the TIs is that besides the conventional semiconductor bulk band 
structure, they also exhibit topological surface conduction states. Some of the best thermoelectric materials are also three-dimensional topological insulators, such as 
$Bi_{2}Te_{3}$, $Bi_{2}Se_{3}$, and $Sb_{2}Te_{3}$ \cite{Xu2017,gooth2018}.
 
Thermoelectric devices must have at least a $ZT > 3$ in order to attain industrial and household spread; this efficiency has been improved over the years, but it was not attained 
until now \cite{Jian17}. This is the reason why thermoelectric generators or thermoelectric refrigerators are not part of our daily technology. They are used in particular fields 
like the satellite and aerospace industry, where the advantages of not having movable parts and not requiring maintenance overshadow the low efficiency \cite{Jian17}. One example 
of this is the radioisotope thermoelectric generator (RTG) \cite{Bourouis20}, a nuclear electric generator that exploits a radioactive atom's natural decay, usually plutonium 
dioxide ${^{238}}PuO_{2}$, converting, via the Seebeck effect, the heat released by the disintegrated atoms into electricity. Furthermore, in the context of today's climate change,
research on new thermoelectric materials that improve thermal efficiency is essential as part of our efforts to obtain environmentally clean sources of energy. 

In this investigation, we focus on studying the semiconductor single-electron transistor (SET), which is the experimental realization of the single impurity Anderson model 
(SIAM) for finite electronic correlation $U$. The SIAM was experimentally realized by the Goldhaber-Gordon group \cite{Goldhaber1998}, with complete control over all of the 
model's parameters. They measured the electric conductance of a SET and showed its universal character. Recently, interest in studying the thermoelectric properties of the 
SET has greatly increased and has given rise to several papers that have discuss it, both from the theoretical side 
\cite{Yoshida2009,Costi2010,Hershfield13,Donsa14,Talbo2017,Costi20191,Costi20192,Thierschmann2019,Eckern2020} and from the experimental one 
\cite{Heremans2004,Scheibner2007,Hoffmann2009,Hartman2018,Artis2018,Dutta2017,Dutta2019}. A useful review can be found in the references \cite{Jian17,Rosa16,Benenti2017}.

Universal relations for the thermal dependence of the thermodynamic properties, in the Kondo regime, for the SIAM are well known, and a didactic discussion can be found in  
Hewson's book \cite{Hewson}.  Costi $et$ $al$. \cite{Costi94,Costi2010}, showed that in the Kondo limit of the SIAM, the thermoelectric transport coefficients (TTC) are only 
functions of $T^{*}=\left(\frac{T}{T_{K}}\right)$, with the temperature $T$ normalized by the Kondo temperature ($T_{K}$). For simplicity, we  only use 
$\left(\frac{T}{T_{K}}\right)$ in the figures, in the text, we use $T^{*}$. They also showed that by employing the NRG, the electric conductance, the temperature normalized 
thermal conductance, and the thermopower exhibit a universal behavior in the Kondo regime. 

The universal behavior of $G\left(T^{*}\right)$ in semiconductor nanostructures was studied in earlier papers \cite{Seridonio1_2009, Seridonio2_2009}. The authors derived an
analytical expression that maps the thermal dependence of $G\left(T^{*}\right)$, of a Kondo-asymmetric condition of the SIAM to a universal conductance function 
$G_{S}\left(T^{*}\right)$. The corresponding TTC  $L_{0}(T^{*})$ (Eq. \ref{Lo}), was calculated by employing the NRG. This analytical mapping is parametrized by $T_{K}$ and the 
ground state phase shift $\delta=\frac{\pi n_{d}}{2}$, which is related to the Friedel's sum rule \cite{Langreth66,Hewson}. For brevity,  we call this quantity parameter 
$\delta$. In this investigation, we derive similar analytical expressions associated with the other TTCs, allowing us to obtain both the thermopower and the thermal conductance in a 
Kondo asymmetric situation, employing only universal functions derived from the thermal dependence of the symmetric SIAM. Again those analytical functions are parametrized by the 
Kondo temperature and the parameter $\delta$.  

In Sec. \ref{sec2}, we define the problem and the model employed in the study of the SET. In Sec. \ref{sec3}, we define the thermoelectric properties and their relation with 
the TTCs. In Sec. \ref{sec4}, we develop the calculations of a mapping between the thermal dependence of the TTCs coefficients in the symmetric limit and the asymmetric 
condition in the quantum dot's Kondo regime. In Sec. \ref{sec5}, we present the results of our research associated with the TTCs and discuss their physical consequences. In 
Sec. \ref{sec6}, we make a prospect of comparison between our results and the available experimental one. In Sec. \ref{sec7}, we present a summary and the conclusions of this 
investigation. In Appendix \ref{method}, we develop a methodology to apply universal TTCs to experimental thermoelectric properties; we obtained simple analytical fitting functions 
that can be used to predict these properties' observed behavior.

\section{The Hamiltonian for the SET} 
\label{sec2}

In this investigation, we extend the previous results obtained 
for the electrical conductance $G\left(T^{*}\right)$, associated with the TTC $L_{0}\left(T^{*}\right)$ 
\cite{Costi2010,Yoshida2009,Seridonio1_2009}, to all the other TTCs: 
$L_{1}\left(T^{*}\right)$ and $L_{2}\left(T^{*}\right)$. We map the thermal dependence of the TTCs for a Kondo-asymmetric situation as a function of the TTCs for a 
Kondo-symmetric condition. These mappings are obtained in terms of the renormalized Kondo temperature $T^{*}$ and the parameter $\delta$.

\begin{figure}[htp]
\centerline{\resizebox{3.3in}{!}{
\includegraphics[clip,width=0.20\textwidth,angle=0.]{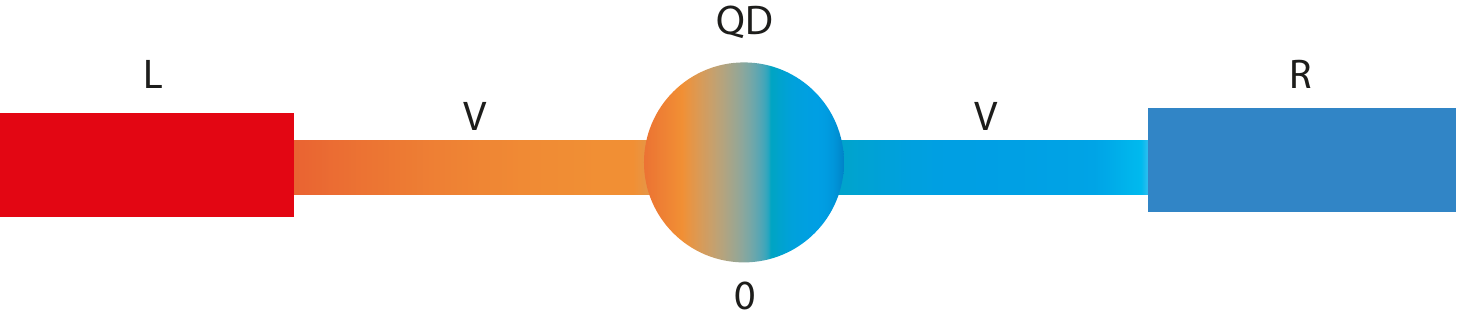}}}
\caption{Schematic picture of a semiconductor electron 
transistor (SET): a quantum dot embedded into conduction leads.}\label{fig01}
\end{figure}

The standard Hamiltonian for the SET studied in this investigation can be written as

$$
H =\sum_{\mathbf{k},\sigma } \sum_{\alpha = L,R} E^{\alpha}_{\mathbf{k},\sigma}
c^{\alpha \dagger}_{\mathbf{k},\sigma}c^{\alpha}_{\mathbf{k},\sigma } 
+\left(E_{d}n_{d}+U n_{d\uparrow} n_{d\downarrow}\right) \nonumber \\
$$
\begin{equation}
+\sum_{\alpha = L,R}\sum_{\mathbf{k},\sigma} \frac{V_{\alpha}}{\sqrt{2N}}\left(c_{d\sigma}^{\dagger}c^{\alpha}_{\mathbf{k},\sigma}+
c^{\alpha \dagger}_{\mathbf{k},\sigma}c_{d\sigma }\right) , 
\label{Eq.1}
\end{equation}
where the first term represents the left ($\alpha=L$) and right 
($\alpha=R$) leads, characterized by hot and cold free conduction electron (c-electrons) reservoirs, respectively. The quantum dot is embedded in the leads as visually 
represented in Fig. \ref{fig01}. The second term describes the QD characterized by the local dot energy $E_{d}$, and $U$ represents the onsite Coulomb repulsion between the 
electrons of the QD \cite{Grobis,Kretinin,Park}. The third term corresponds to the tunneling between the immersed dot and the left (L) and right (R) semi-infinite leads. The 
amplitude $V_{\alpha}$ is responsible for the tunneling between the QD and the lead $\alpha$. For simplicity, we assume symmetric junctions (i.e. $V_{\alpha}=V_{L}=V_{R}=V$) and 
identical leads (i.e. $E^{L}_{\mathbf{k},\sigma}=E^{R}_{\mathbf{k},\sigma}=E_{\mathbf{k},\sigma}$) connecting the QD to the quantum wire. The $L(R)$ semi-infinite leads 
comprise $N$ states $c^{L}_{\mathbf{k},\sigma}$ ($c^{R}_{\mathbf{k},\sigma}$) with energies defined by the linear dispersion relation 
$E_{\mathbf{k},\sigma}=\left(k-k_{F}\right)v_{F}$ ($0\leq k \leq 2k_{F}$) so that the bandwidth is $2D=2v_{F}k_{F}$, with $D$ being the half-width of the conduction band. In 
all the numerical calculations, we consider the unit of energy to be $D=1$.

\section{Thermoelectric properties}
\label{sec3}

We calculated the electrical conductance $G(T)$, the thermal conductance $\kappa_{e}(T)$ and the thermopower $S(T)$ (Seebeck effect) in terms of the transport coefficients, 
following  the standard textbook derivation \cite{Mahan,Ziman}, and the results are

\begin{equation}
\label{G}
G(T)=e^{2}L_{o}(T),
\end{equation}
\noindent 
\begin{equation}
\label{K}
\kappa_{e}(T)=\frac{1}{T}\left(L_{2}(T)-\frac{L_{1}^{2}(T)}{L_{o}(T)}\right), 
\end{equation}
and 
\begin{equation}
S=\left(\frac{-1}{eT}\right) \frac{L_{1}(T)}{L_{o}(T)} .
\label{S}
\end{equation}

To calculate the transport coefficients $L_{o}(T)$, $L_{1}(T)$, and $L_{2}(T)$, we employed the results derived by Dong and X. L. Lei \cite{Dong_02}. They considered the particle 
current and thermal flux formulas, through an interacting QD connected to the leads at different temperatures, within the Keldysh non-equilibrium Green's functions (GF) formalism. 
The electric and thermoelectric transport coefficients were obtained in the presence of the chemical potential and temperature gradients, 
with the Onsager relation in the linear regime being automatically satisfied. The transport coefficients are given by

\begin{equation}
L_{n}(T)=\frac{2}{h}\int{\left(-\frac{\partial
n_{F} (\omega,T) }{\partial \omega} \right) \omega^{n}\tau(\omega,T) d\omega} ,
\label{L11} 
\end{equation}
where $n_{F}(\epsilon,T)=1/(1+e^{(\epsilon-\mu)/k_{B}T})$ is the Fermi-Dirac distribution, with $\mu$ being the chemical potential  and the transmittance  $\tau(\omega,T)$ 
is given by 
\begin{equation}
\tau(\omega,T)= \pi \Gamma \rho_{d}(\omega,T),
\label{Trans1} 
\end{equation}
 where $\rho_{d}(\omega,T)$ is the spectral density of the QD and $\Gamma=\pi\rho_{c}(\mu=0)V^{2}$ is the Anderson parameter, which is a measure of the $d-$level width, and 
$\rho_{c}(\omega)=\frac{1}{2D}$ is the flat conduction density of states of the leads.

A useful quantity that indicates the system performance is the thermoelectric dimensionless figure of merit $ZT$ \cite{Joffe_1959}, which is given by 
\begin{equation}
ZT=\frac{S^{2}TG}{\kappa_{e}} .
\label{ZT}
\end{equation}

\section{Universal mapping: Transmission coefficient and thermoelectric coefficients}
\label{sec4}

Following the reference \cite{Seridonio1_2009}, we introduce the normalized even ($a_{\mathbf{k},\sigma}$) and odd ($b_{\mathbf{k},\sigma}$) operators, in order  to exploit the 
inversion symmetry of the system

\begin{equation}
 a_{\mathbf{k},\sigma}=\frac{1}{\sqrt{2}}\left(c^{L}_{\mathbf{k},\sigma}+c^{R}_{\mathbf{k},\sigma}\right),
\label{ak}
\end{equation}
\begin{equation}
 b_{\mathbf{k},\sigma}=\frac{1}{\sqrt{2}}\left(c^{L}_{\mathbf{k},\sigma}-c^{R}_{\mathbf{k},\sigma}\right).
\label{bk}
\end{equation}

It is convenient to write the Hamiltonian of Eq. \ref{Eq.1} on the basis of the new operators ($a_{\mathbf{k},\sigma}$ and $b_{\mathbf{k},\sigma}$), to  ``split'' it into two 
decoupled pieces $H=H_{A}+H_{B}$, with

\begin{eqnarray}
H_{A}&=&\sum_{\mathbf{k},\sigma}E_{\mathbf{k},\sigma}a_{\mathbf{k},\sigma}^{\dagger}a_{\mathbf{k},\sigma}+
\sum_{\sigma}V\left(f_{0}^{\dagger}c_{d,\sigma}+h.c\right) \\ \nonumber 
&& + \left(E_{d}n_{d}+U n_{d\uparrow} n_{d\downarrow}\right),
\label{HA}
\end{eqnarray}

where $f_{0}=\sum_{\mathbf{k},\sigma}\left(\frac{a_{\mathbf{k},\sigma}}{\sqrt{N}}\right)$, is the traditional NRG shorthand notation, and $H_{B}$ is given by

\begin{equation}
 H_{B}=\sum_{\mathbf{k},\sigma}E_{\mathbf{k},\sigma}b_{\mathbf{k},\sigma}^{\dagger}b_{\mathbf{k},\sigma}.
\label{HB}
\end{equation}

The Hamiltonian $H_{B}$ is quadratic and can be exactly diagonalized and decoupled from the quantum dot. On the contrary, the Hamiltonian $H_{A}$ ``carries'' all the correlation
effects of the quantum dot and the coupling between it and the conduction band. Due to this, the Hamiltonian $H_{A}$ is the only one relevant for obtaining the transmittance and 
the spectral density of states for the quantum dot, and can be written as \cite{Seridonio1_2009}

\begin{eqnarray}
\rho_{d}(\omega,T)&=&\frac{1}{f(\omega,T)}\sum_{mn,\sigma}\frac{e^{-\beta E_{m}}}{Z(T)}\\ \nonumber
&&  |<n|c_{d,\sigma}^{\dagger}|m>|^{2}\delta(E_{mn}-\hbar\omega).
\label{denQD}
\end{eqnarray}

Here $|m>$ and $|n>$ are the eigenstates of $H_{A}$, with eigenvalues $E_{m}$ and $E_{n}$, respectively, $E_{mn}=E_{m}-E_{n}$, and $Z(T)$ is the partition function for 
the Hamiltonian $H_{A}$. The Hamiltonian $H_{B}$ is not dependent on $c_{d,\sigma}$, $\left[H_{B},c_{d,\sigma}\right]=0$, and only the eigenvalues and eigenvectors of $H_{A}$ are 
required  to obtain $\rho_{d}(\omega,T)$. On the other hand, to calculate the matrix elements $<n|c_{d,\sigma}^{\dagger}|m>$ in Eq. $16$, we evaluate the commutator 
$\left[H_{A},a_{\mathbf{q},\sigma}\right]$

\begin{equation}
\left[H_{A},a_{\mathbf{q},\sigma}\right]=E_{q,\sigma}a_{\mathbf{q},\sigma}^{\dagger}+\frac{V}{\sqrt{N}}c_{d,\sigma},
\label{ComA}
\end{equation}
and performing the summation over $q$ and $\sigma$ we obtain
\begin{equation}
\left[H_{A},f_{0}^{\dagger}\right]=\frac{1}{\sqrt{3}}f_{1}^{\dagger}+Vc_{d}^{\dagger},
\label{ComA-2}
\end{equation}
where 
\begin{equation}
f_{1}=\sqrt{\frac{3}{N}}\sum_{q}\left(\frac{E_{q}}{D}\right)a_{q},
\label{f1} 
\end{equation}
define a new NRG shorthand notation operator.

Equation \ref{ComA-2} permits to relate \hspace{0.1cm} the matrix elements $<n|c_{d,\sigma}^{\dagger}|m>$ in Eq. $16$ with the same matrix elements of the operators $f_{0}$ 
and $f_{1}$,

\begin{equation}
 V<m|c_{d}^{\dagger}|n>=E_{mn}<m|f_{0}^{\dagger}|n>-\sqrt{3}D<m|f_{1}^{\dagger}|n> ,
\label{Matrix1}
\end{equation}
and a Schrieffer-Wolff transformation of the Hamiltonian $H_{A}$, allows us to write it in the Kondo form \cite{Seridonio1_2009}

\begin{equation}
 H_{J}=\sum_{k}E_{l}g_{l}^{\dagger}g_{l}+J_{W}\sum_{\mu,\nu}\Phi_{0\mu}^{\dagger}\mathbf{\sigma}_{\mu\nu}\Phi_{0\nu} \cdot \mathbf{S},
\label{HA-Kondo}
\end{equation}
where the $g_{l}$ operators are the eigenoperators of the fixed-point Hamiltonian, associated with the unstable local moment (LM) condition of the Anderson impurity model
(see ref. \cite{Seridonio1_2009} for details). Here, $J_{W}=4D \frac{\Gamma U}{\pi|E_{d}|\left(E_{d}+U\right)}cos^{2}\delta_{LM}$, where $\delta_{LM}$ is the quantum scattering 
phase shift, associated with the LM fixed point, and

\begin{equation}
\Phi_{0}=\frac{1}{\sqrt{N}}\sum_{l}g_{l} ,
\label{Phi0}
\end{equation}
where in the symmetric condition $\delta_{LM}=0$ and $\Phi_{0}=f_{0}$. 

The second term in the Eq. \ref{HA-Kondo} is responsible, in the Kondo regime, for the evolution from the LM fixed point to a 
Fermi liquid (FL) fixed point, associated with an antiferromagnetic 
$J_{W}$ coupling, characteristic of the Kondo effect. We can define the operator 

\begin{equation}
 \Phi_{1}=\sqrt{\frac{3}{N}}\sum_{l}\left(\frac{E_{l}}{D}\right)g_{l} ,
\label{Phi1}
\end{equation}
in an way analogous to the $f_{1}$ operator's definition (Eq. \ref{f1}). In the symmetric condition $\Phi_{1}=f_{1}$, something similar happens with $\Phi_{0}=f_{0}$.

Eqs. \ref{ComA} and \ref{Matrix1} show the universal character of the product $V<m|c_{d}^{\dagger}|n>$ in the symmetric point (remember $\Phi_{0}=f_{0}$ and $\Phi_{1}=f_{1}$ in this 
condition). In order to explore what happens in the asymmetric condition, it is necessary to relate the operators $f_{0}$ and $f_{1}$ to $\Phi_{0}$ and $\Phi_{1}$,  (see Appendix A2 
of reference \cite{Costi94}). Substituting Eq. $(A21)$ of the reference \cite{Costi94} in Eq. \ref{Matrix1} we obtain

\begin{equation}
 \sqrt{\pi\rho\Gamma}<m|c_{d}^{\dagger}|n>=\alpha_{0}<m|\Phi_{0}^{\dagger}|n>+\alpha_{1}<m|\Phi_{1}^{\dagger}|n>.
\label{Matrix2}
\end{equation}

Performing the substitution of Eq. \ref{Matrix2} in Eq. \ref{ComA}, we obtain the localized QD spectral density $\rho_{d}(\omega,T)$, which can be written as

\begin{equation}
\pi\rho\Gamma\rho_{d}(\omega,T)=\alpha_{0}^{2}\rho_{0}(\omega,T)+\alpha_{1}^{2}\rho_{1}(\omega,T)+\alpha_{0}\alpha_{1}\rho_{(01)}(\omega,T),
\label{densid}
\end{equation}
where $\rho_{0}(\omega,T)$, $\rho_{1}(\omega,T)$ and $\rho_{(01)}(\omega,T)$ are universal expressions of the Kondo regime and are given by
\begin{eqnarray}
\rho_{j}(\omega,T)&=&\sum_{mn}\frac{e^{-\beta E_{m}}}{Z(T)f(\omega,T)}|<n|\Phi_{j}|m>|^{2} \times \\ \nonumber
&&  \delta(E_{mn}-\hbar\omega), \hspace{1.0cm} (j=0,1) ,
\label{rho}
\end{eqnarray}
and 
\begin{eqnarray}
\rho_{(01)}(\omega,T)&=&\sum_{mn}\frac{e^{-\beta E_{m}}}{Z(T)f(\omega,T)}(<n|\Phi_{0}|m> \times \\ \nonumber 
&& <n|\Phi_{1}|m> + c.c.)  \delta(E_{mn}-\hbar\omega).
\label{rho01}
\end{eqnarray}

Substituting Eq. \ref{densid} in Eq. \ref{Trans1}, the transmittance at energy $\epsilon=\frac{h\omega}{2\pi}$ and temperature $T$ is given 
by
\begin{equation}
\label{Trans}
\tau(\omega,T)=
\frac{\alpha_{0}^{2}}{\rho}\rho_{0}(\omega,T)+
+\frac{\alpha_{1}^{2}}{\rho}\rho_{1}(\omega,T)+\frac{\alpha_{0}\alpha_{1}}{\rho}\rho_{(01)}(\omega,T).
\end{equation}

The universal expressions $\rho_{0}(\omega,T)$, $\rho_{1}(\omega,T)$ and $\rho_{(01)}(\omega,T)$ ``carry'' the thermal dependence of $\tau(\omega,T)$. The important point that 
should be stressed here is that the transmittance is the key physical quantity that enters the calculations of all the thermoelectric coefficients given by Eq. \ref{L11}. All 
the dependence of the parameters of the model is taken into account through the coefficients $\alpha_{0}$ and $\alpha_{1}$, given by the reference \cite{Seridonio1_2009}

\begin{equation}
\alpha_{0}^{2}=cos^{2}(\delta) ,
\label{alpha0}
\end{equation}
\begin{equation}
\alpha_{1}^{2}=\frac{6}{\pi^{2}}sin^{2}(\delta) .
\label{alpha1} 
\end{equation}

Taking into account the result of Eqs. \ref{L11} and \ref{Trans}, it is possible to compute the thermal dependence of the linear TTCs  as a function of ($T^{*}$)

\begin{equation}
L_{0}\left(T^{*}\right)-\frac{1}{h}=-\left(L_{0}^{S}\left(T^{*}\right)-\frac{1}{h}\right)cos(2\delta),
\label{Lo}  
\end{equation}
where $L_{0}^{S}\left(T^{*}\right)$ is the universal coefficient $L_{0}$ in the electron-hole symmetric condition of the model, 
when $E_{d}=\frac{-U}{2}$ and $\delta=\frac{\pi n_{d}}{2}$, with $n_{d}=1$. All the thermal dependence of $L_{0}\left(T^{*}\right)$ is contained in the universal 
function $L_{0}^{S}\left(T^{*}\right)$. The function $cos(2\delta)$ carries all the parameter dependence apart from 
temperature $T$, and is characteristic of the asymmetric conditions for the model ($\delta\neq\frac{\pi}{2}$).

Taking into account that ${\cal{G}}_{2}(T)=e^{2}L_{0}(T)$, we can write the Eq. \ref{Lo} in the same form as a result previously obtained in reference \cite{Seridonio1_2009} 

\begin{equation}
G_{2}\left(T^{*}\right)-{\cal{G}}_{2}=
-\left({\cal{G}}_{2}^{S}\left(T^{*}\right)-
{\cal{G}}_{2}\right)cos(2\delta),
\label{G_Seridonio}  
\end{equation}
where we also should observe that the $\rho_{(01)}(\omega,T)$ term in 
Eq. \ref{Trans}, due to particle-hole symmetry arguments, makes no contribution to the thermoelectric coefficients $L_{0}$ \cite{Seridonio_2009} and $L_{2}$, but contributes 
to $L_{1}$, as indicated in Eqs. \ref{L1} and \ref{L01}.

The evaluation of the $L_{1}$ coefficient, employing the result for the transmission coefficient (Eq. \ref{Trans}) in Eq. \ref{L11} 
($n=1$), and taking into account the parity condition of the integrand, give us
\begin{equation}
L_{1}\left(T^{*}\right)=L_{(01)}\left(T^{*}\right)cos(\delta)sin(\delta) , 
\label{L1}
\end{equation}
where 
\begin{equation}
L_{(01)}(T)=\frac{\sqrt{2}\pi}{h\rho}\int_{-D}^{D}{\epsilon \rho_{(01)}(\epsilon,T)\left(-\frac{\partial f(\epsilon,T)}{\partial \epsilon}\right) d\epsilon} ,
\label{L01}
\end{equation}
which is an universal function of $\left(T^{*}\right)$ in the symmetric Kondo condition, and contains all the thermal dependence of $L_{1}\left(T^{*}\right)$. 

Finally, employing the result for the transmittance (Eq.\ref{Trans}) in 
Eq. \ref{L11} ($n=2$), we obtain the 
$L_{2}\left(T^{*}\right)$ coefficient. Again, we take into account the parity of the integrand 
 
\begin{equation} 
\frac{L_{2}\left(T^{*}\right)}{\left(\frac{k_{B}T}{T_{K}}\right)^{2}}-\frac{\pi^{2}}{6}=-cos(2\delta)\left(\frac{L_{2}^{S}\left(T^{*}\right)}
{\left(\frac{k_{B}T}{T_{K}}\right)^{2}}-\frac{\pi^{2}}{6}\right).
\label{L2}
\end{equation}

The quantity $\frac{L_{2}^{S}\left(T^{*}\right)}{\left(\frac{k_{B}T}{T_{k}}\right)^{2}}$ is a universal function of $\left(T^{*}\right)$, obtained 
in terms of $L_{2}^{S}(T)$, the coefficient for the symmetric condition of the model. As in the previous cases, 
all the thermal dependence of Eq. \ref{L2} in any asymmetric condition of the model is contained in 
$\frac{L_{2}^{S}\left(T^{*}\right)}{\left(\frac{k_{B}T}{T_{K}}\right)^{2}}$, and all the dependence on the parameters of the model is taken into account through the scattering
 phase shift factor $\delta$.

\section{Results and Discussion: Universal Mapping}
\label{sec5}

In previous papers, \cite{Seridonio_2009,Seridonio1_2009,Seridonio2_2009,Seridonio_2010} one of us (L. N. Oliveira), argued that it is possible to employ experimental data 
of electrical conductance $G\left(T^{*}\right)$ to obtain the parameter $\delta$, and with it "check" the validity of the Eq. \ref{Lo} for the SET. Computations for the case 
of a side-coupled quantum dot were also considered in references \cite{Seridonio_2009,Seridonio2_2009,Seridonio_2010}, including a comparison with experimental 
results \cite{Seridonio_2009}. In all those previous papers, the numerical calculations were done employing NRG, including the computation of the parameter $\delta$. 

The NRG logarithmic discretization parameter employed in the simulations of this work was $\Lambda=2.25$ and the chemical potential, $\mu=0.0D$. The Kondo temperature,
$T_{K}$, for each case was obtained by computing the value of the temperature, where the electrical conductance attains the value
$G(T_{K})=\frac{G_{o}}{2}=\frac{e^{2}}{h}$ \cite{Goldhaber1998PRL}.

\begin{figure}[htp] 
\begin{center}
\includegraphics[clip,width=0.45\textwidth,angle=0.0]{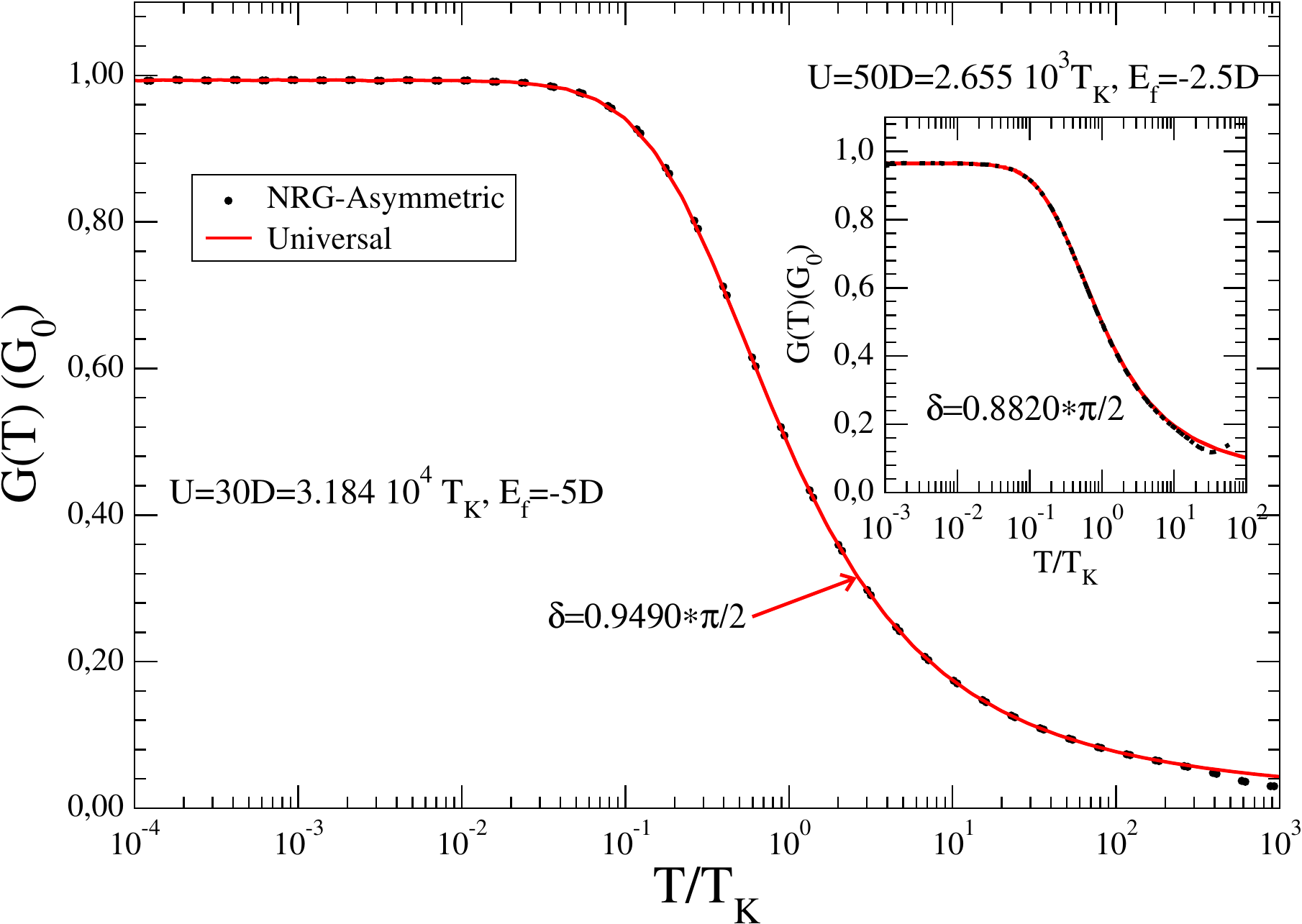}
\caption {(Color online) The electrical conductance $G(T)$ vs. 
$T^{*}$, corresponding to the  Kondo regime. In the inset we show another situation, closer to the crossover between the Kondo and intermediate valence regime.}
\label{fig1}
\end{center}
\end{figure}
Fig. \ref{fig1} shows the results obtained for the electrical conductance $G(T)$ vs. $T^{*}$, in units of 
$G_{o}=\frac{2e^{2}}{h}$, in an asymmetric situation employing the results of the symmetrical limit as 
shown in Fig. \ref{G-Uni} of Appendix \ref{method}. In the main panel, we plot results corresponding to the Kondo regime, employing the following parameters: $E_{f}=-5.0D$,
$U=30.0D$, with the Kondo temperature being $T_{K}=9.422.10^{-4}D$. The agreement between the calculated NRG asymmetric results and those obtained employing the NRG symmetric 
one,  the Eq. \ref{Lo}, is notable. The parameter $\delta$ computed by the NRG for the asymmetric case is $\delta=0.9490\frac{\pi}{2}$, which confirms the validity of 
Eq. \ref{Lo} for the SET, previously obtained in reference \cite{Seridonio1_2009}. In the inset, we plot a situation closer to the crossover transition 
between the Kondo and intermediate valence regime, with $U=50.0D$, $E_{f}=-2.5D$, $T_{K}=1.883 10^{-2}D$, and $\delta=0.882\frac{\pi}{2}$. The agreement obtained is notable 
for temperatures below $T \simeq 10 T_{K}$, but for temperatures above this value, the results show a small departure from each other, due to the rising of charge 
fluctuations not being well described by the present treatment. 

\begin{figure}[htp] 
\begin{center}
\includegraphics[clip,width=0.45\textwidth,angle=0.0]{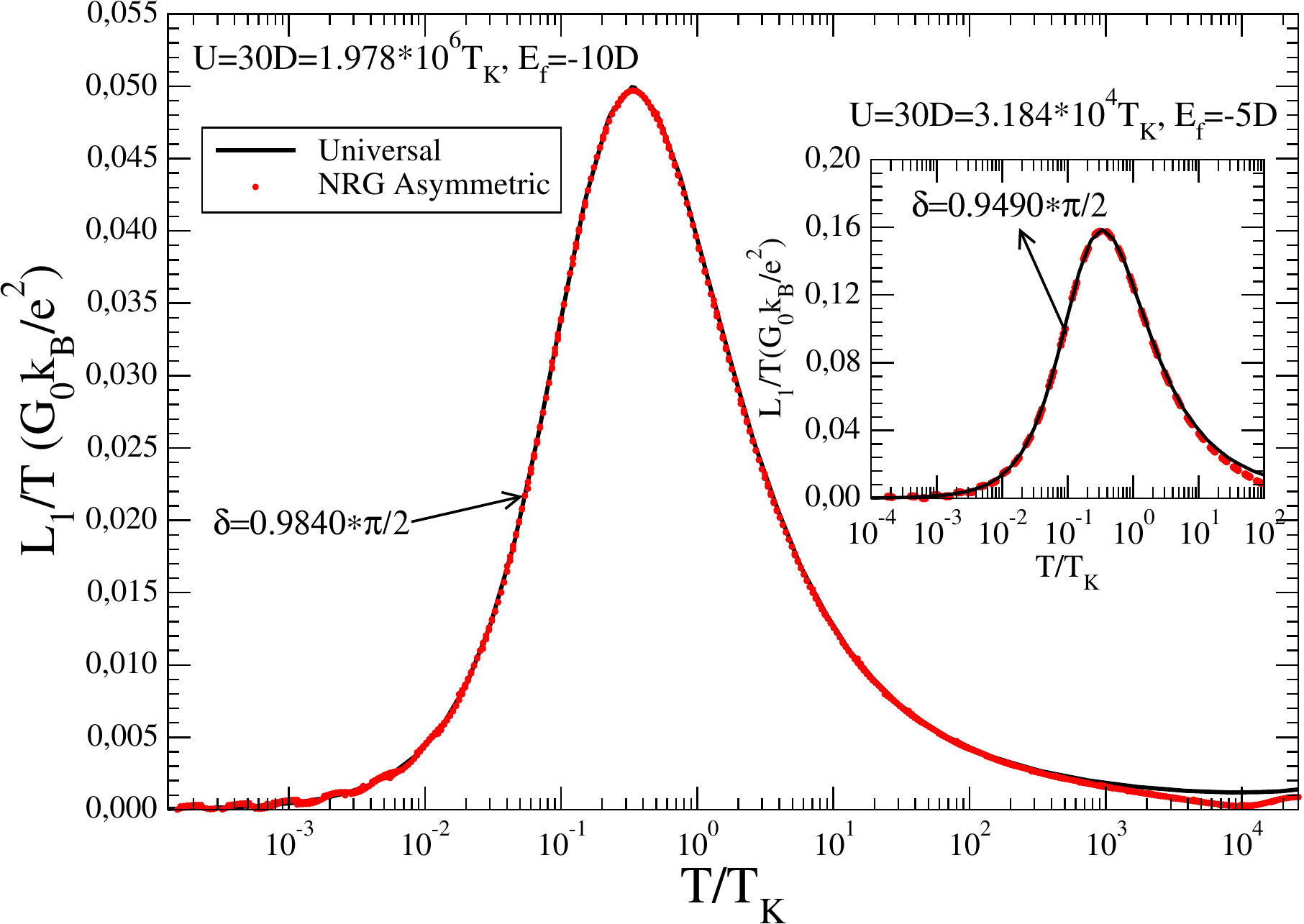}
\caption{(Color online) Universal thermoelectric coefficient  $\left(\frac{L_{1}}{T}\right)$, expressed in $\frac{G_{o}k_{B}}{e^{2}}$ units, vs. $T^{*}$ for the 
asymmetric Kondo limit. In the inset, we show the same results for an asymmetric case, in the crossover from an intermediate valence situation to the Kondo limit.}
		\label{FitL1}
   \end{center}		
  \end{figure}
In Fig. \ref{FitL1}, in the main panel, we show the results for  
$\left(\frac{L_{1}}{T}\right)$  vs. $T^{*}$ for the asymmetric Kondo limit, with $E_{QD}=-10.0D$, $U=30.0D$ and 
$T_{K}=1.517.10^{-5}D$. The direct NRG computations are shown by the continuous red line, whereas  the results obtained when employing the NRG calculations for the 
particle-hole symmetric case of the SIAM and Eq. \ref{L1} are shown by the black curve. Again, the parameter 
$\delta$ was computed following the same procedure described in the Appendix \ref{method}. We obtained excellent agreement for a large range of temperatures, between 
($10^{-4} T_{K}\leq T \leq 10^{3}T_{K}$). In the inset, we show the same results, but now for a set of parameters closer to the crossover transition region, between 
the Kondo and the intermediate valence regimes, with $E_{QD}=-5.0D$, $U=30.0D$, and $T_{K}=9.422.10^{-4}D$.

\begin{figure}[htp] 
\begin{center}
\includegraphics[clip,width=0.45\textwidth,angle=0.0]{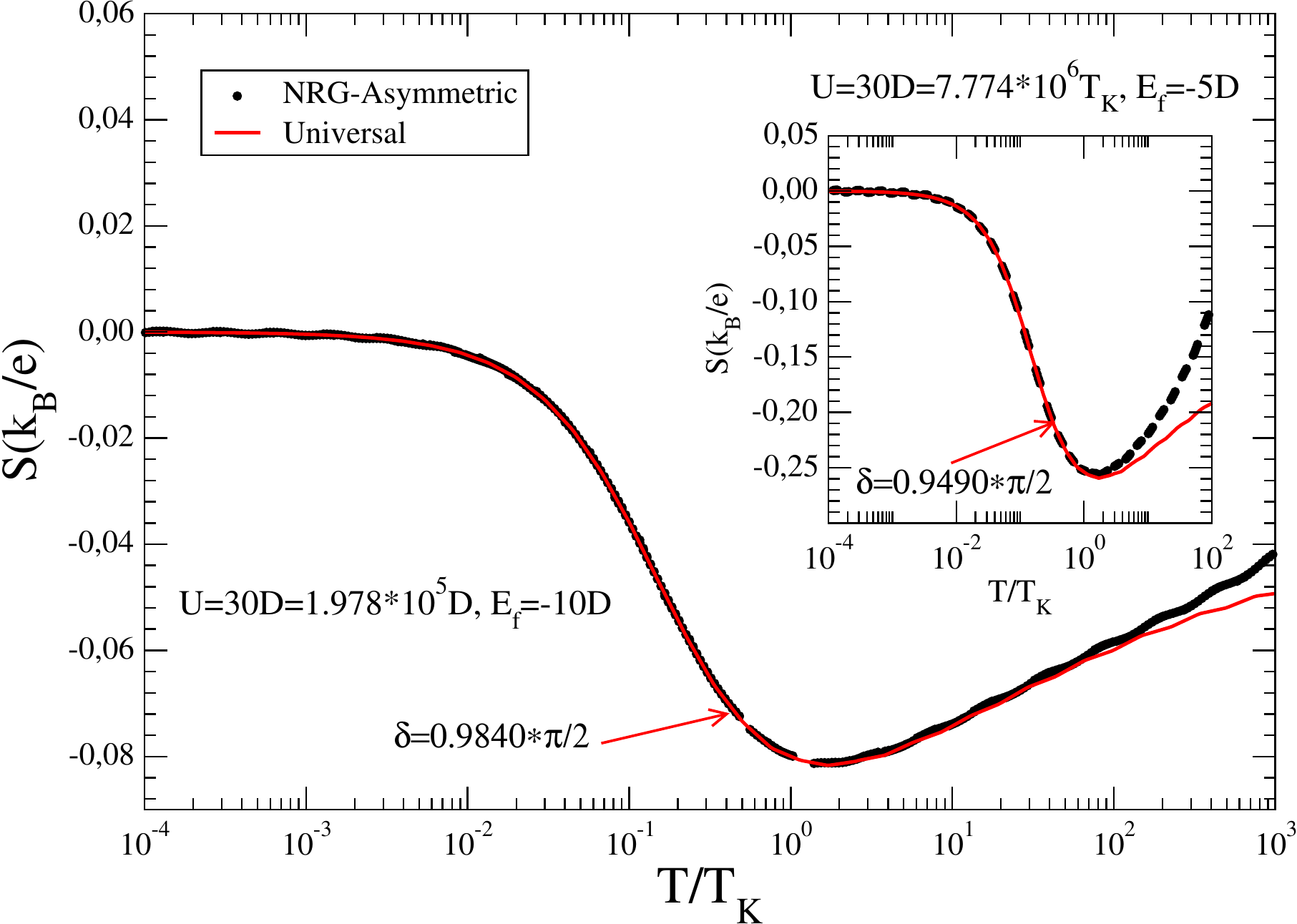}
\caption {(Color online) Thermopower $S(T^{*})$, in $\frac{K_{B}}{e}$ units vs. $T^{*}$. The inset shows the crossover's results from the intermediate valence 
to the Kondo regime.}
\label{Thermopower}
\end{center}
\end{figure}

In Fig. \ref{Thermopower}, we plot the thermopower $S(T^{*})$ vs. 
$T^{*}$ for $E_{QD}=-10.0D$, $U=30.0D$, and $T_{K}=1.517.10^{-5}D$. Employing Eqs. \ref{S}, \ref{Lo}, and \ref{L1} (red curve), we obtain an excellent agreement between
the asymmetric direct NRG results (black curve) and the one employing the symmetric universal TTCs. The minimum at the Kondo temperature manifests the Kondo effect in 
the thermopower $S(T)$ \cite{Costi2010}. There is excellent agreement between both curves up to $T\leq 10^{2}T_{K}$, when charge fluctuations dominate the process. In 
the inset, we represent a crossover from intermediate valence to the Kondo regime $E_{QD}=-5.0D$, $U=30.0D$, and $T_{K}=9.422 10^{-4}D$. Below $T\geq T_{K}$, the
agreement between the two curves is excellent, but above $T_{K}$, there is a visible difference between the two results at higher temperatures. We attribute this difference 
to the intermediate valence region's proximity, because the present treatment does not describe charge fluctuations well.

\begin{figure}[htp] 
\begin{center}
\includegraphics[clip,width=0.40\textwidth,angle=0.0]{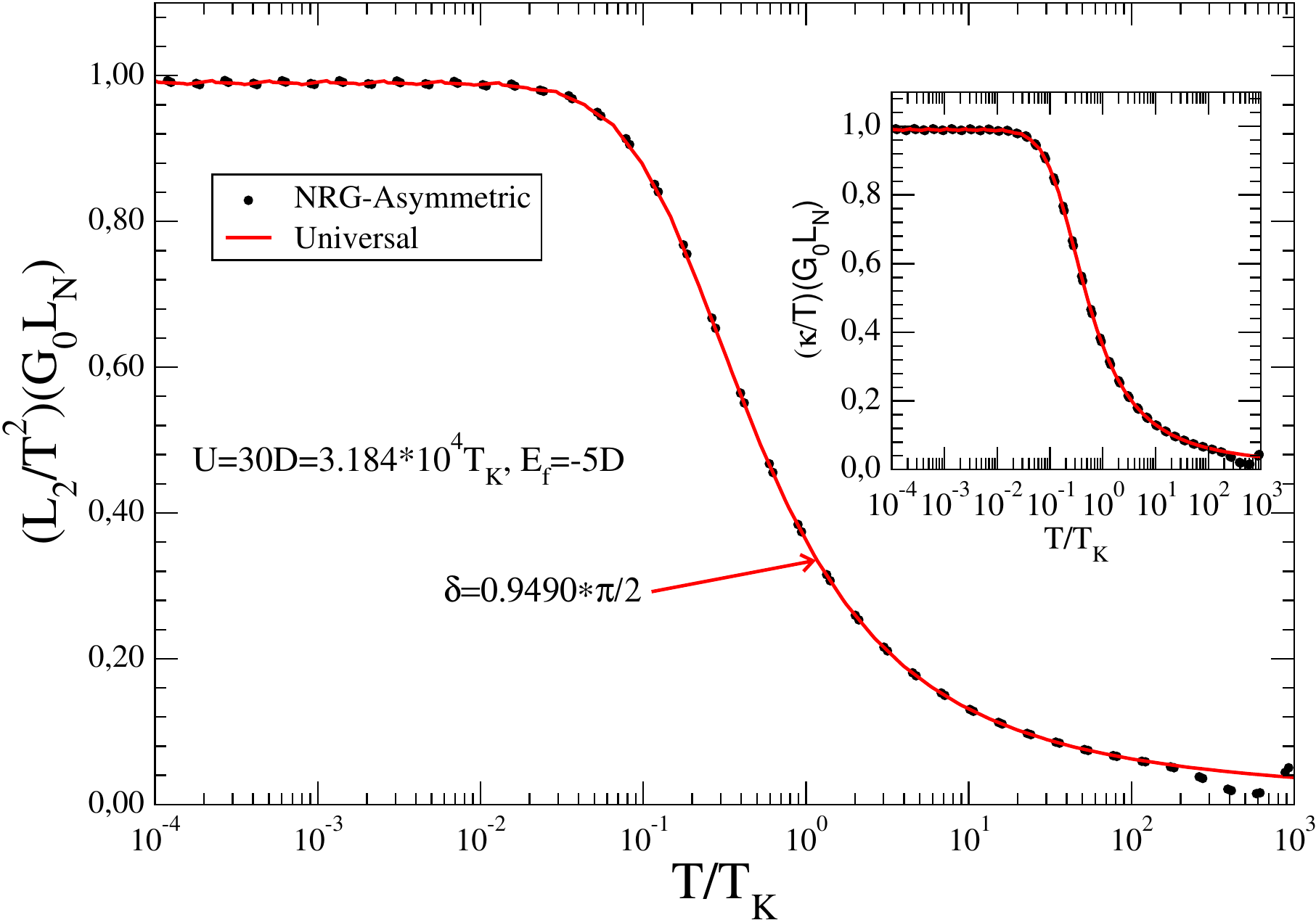}
\caption {(Color online) Universal thermoelectric coefficient 
$\left(\frac{L_{2}}{T^{2}}\right)$, in $G_{o}L_{N}$ units, vs. $T^{*}$ for the asymmetric Kondo limit. The inset shows the 
temperature-normalized electronic contribution to the thermal conductance $\left(\frac{\kappa}{T}\right)$ vs. $T^{*}$.}
\label{fig6}
\end{center}
\end{figure}

In Fig. \ref{fig6}, we plot the results for the universal thermoelectric coefficient for the asymmetric Kondo 
limit, $\left(\frac{L_{2}}{T^{2}}\right)$ vs. $T^{*}$ in $G_{o}L_{N}$ units,  where $L_{N}=(\frac{\pi^{2}}{3})(\frac{k_{B}}{e})^{2}$ is the Lorenz number, 
with $E_{QD}=-5.0D$, $U=30.0D$, and $T_{K}=9.422.10^{-4}D$. Again the agreement obtained between the direct asymmetric NRG results and those achieved employing  
Eq. \ref{L2} and particle-hole symmetric NRG results (fitting presented in Fig. \ref{L2-Uni}) is excellent. In the inset, we show the temperature-normalized electronic 
contribution to the thermal conductance $\left(\frac{\kappa}{T}\right)$ vs. $T^{*}$.  In this case, some small differences appear above $T\geq 30.0 T_{K}$, which is a 
manifestation of the charge fluctuation process, present in this range of temperatures. 

\begin{figure}[htp] 
\begin{center}
\includegraphics[clip,width=0.40\textwidth,angle=0.0]{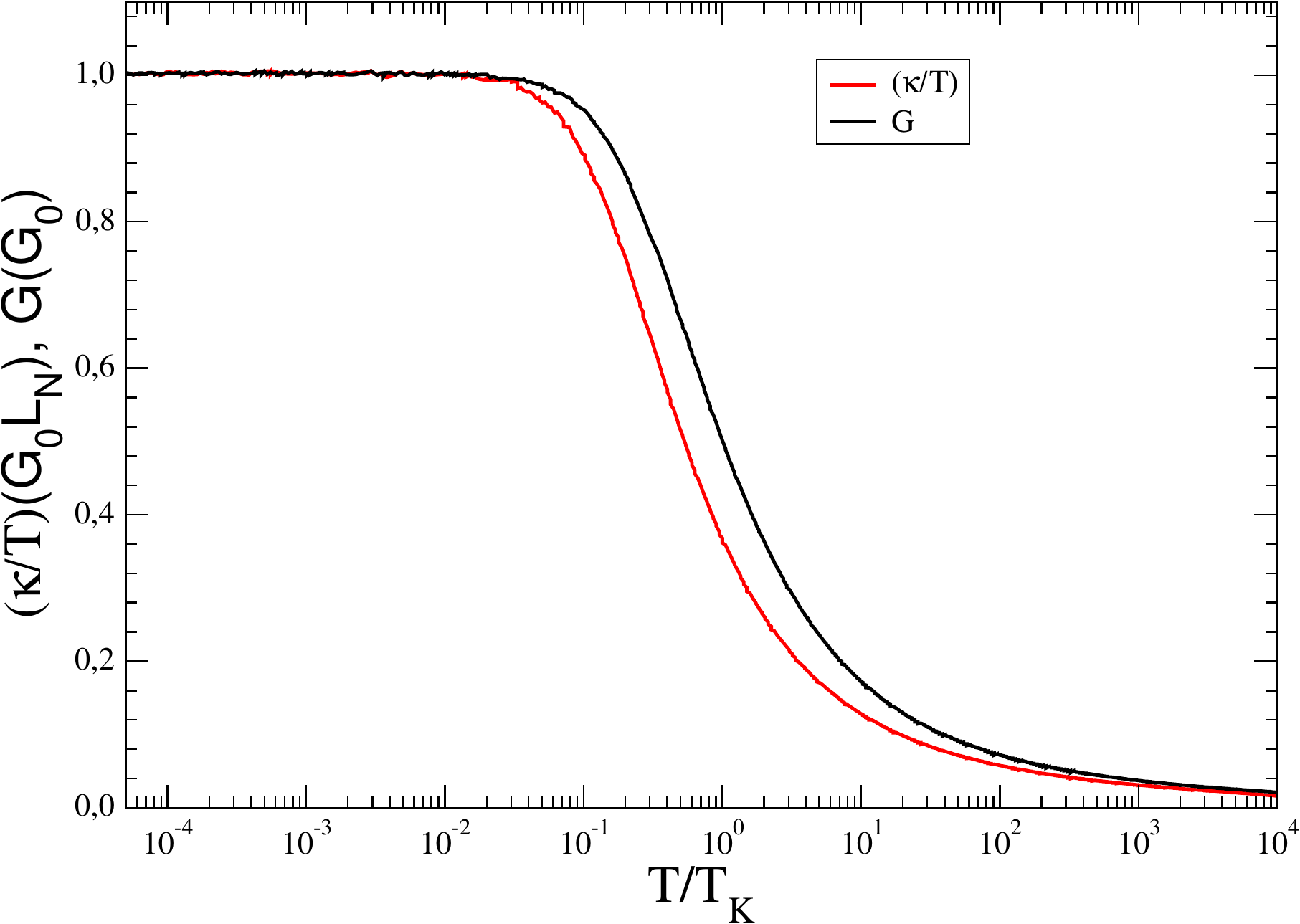}
\caption {(Color online) Thermal dependence of universal quantities in the symmetric limit. Temperature normalized electronic contribution to the thermal conductance, 
$\left(\frac{\kappa(T)}{T}\right)(G_{o}L_{N}$ and the electric conductance $G(T)$ vs. $T^{*}$.}
\label{fig2}
\end{center}
\end{figure}
Fig. \ref{fig2} shows the thermal dependence of the universal quantities in the symmetric limit of the SIAM, employing the following parameters: $U=30.0D$, $E_{f}=-15.0D$. We 
plot the temperature normalized electronic contribution to the thermal conductance, $\left(\frac{\kappa(T)}{T}\right)$ and the electric conductance $G(T)$ vs. 
$T^{*}$. The striking similarity of both curves at low temperatures is associated with the Fermi-liquid character of the system and the validity of the Wiedemann-Franz law
in this temperature range \cite{Costi2010, Yoshida2009}, which leads to the relation $\frac{\kappa(T)}{T}=G(T)$. However, besides the relative closeness of the curves well
below and above the Kondo temperature, the two properties are not equal, once the Kondo temperature rules the electrical conductance, whereas the thermal conductance is 
ruled by a different Kondo scale $T_{K}^{\theta}$, as defined in reference \cite{Costi2010}
\begin{equation}
\frac{\kappa(T=T_{K}^{\theta})}{T_{K}^{\theta}}=\frac{\tilde{\alpha}}{2},
\label{TKtheta}
\end{equation}
where 
\begin{equation}
\tilde{\alpha}=\lim_{T \rightarrow 0} \frac{\kappa(T)}{T} .
\label{const}
\end{equation}

\section{Comparison with experimental results}
\label{sec6}

In this section, we discuss how to use the methodology employing the universal TTCs to calculate the thermoelectric properties from experimental measurements. In  
Appendix \ref{method}, we present a discussion and some examples of applying the universal TTCs methodology to experimental thermoelectric data. 

Unfortunately, we did not find experimental SET works in the literature that performed measurements of the electric and thermal conductances and the thermopower in 
a broad range of temperatures. On the other hand, several papers measured the gate dependence $V_{gate}=V$ of some of these properties for a fixed temperature, $T$ 
\cite{Scheibner2005,Hoffmann2009,Svensson2013,Artis2018,Dutta2019}. We focus on applying the universal TTC methodology to the experimental results of the Artis $et$ $al.$ 
paper \cite{Artis2018}, because they performed several high-quality thermoelectric measurements of Kondo correlated quantum dots (QDs), both below and above the Kondo 
temperature. They measured the electric conductance $G(T)$, thermocurrent $I_{Th}$ normalized by $\Delta T$ (under closed-circuit conditions), and the thermovoltage 
$V_{th}$ (under open-circuit conditions) as a function of the gate voltage $V$ for a fixed temperature.

\begin{figure}[htp] 
\begin{center}
\includegraphics[clip,width=0.40\textwidth,angle=0.0]{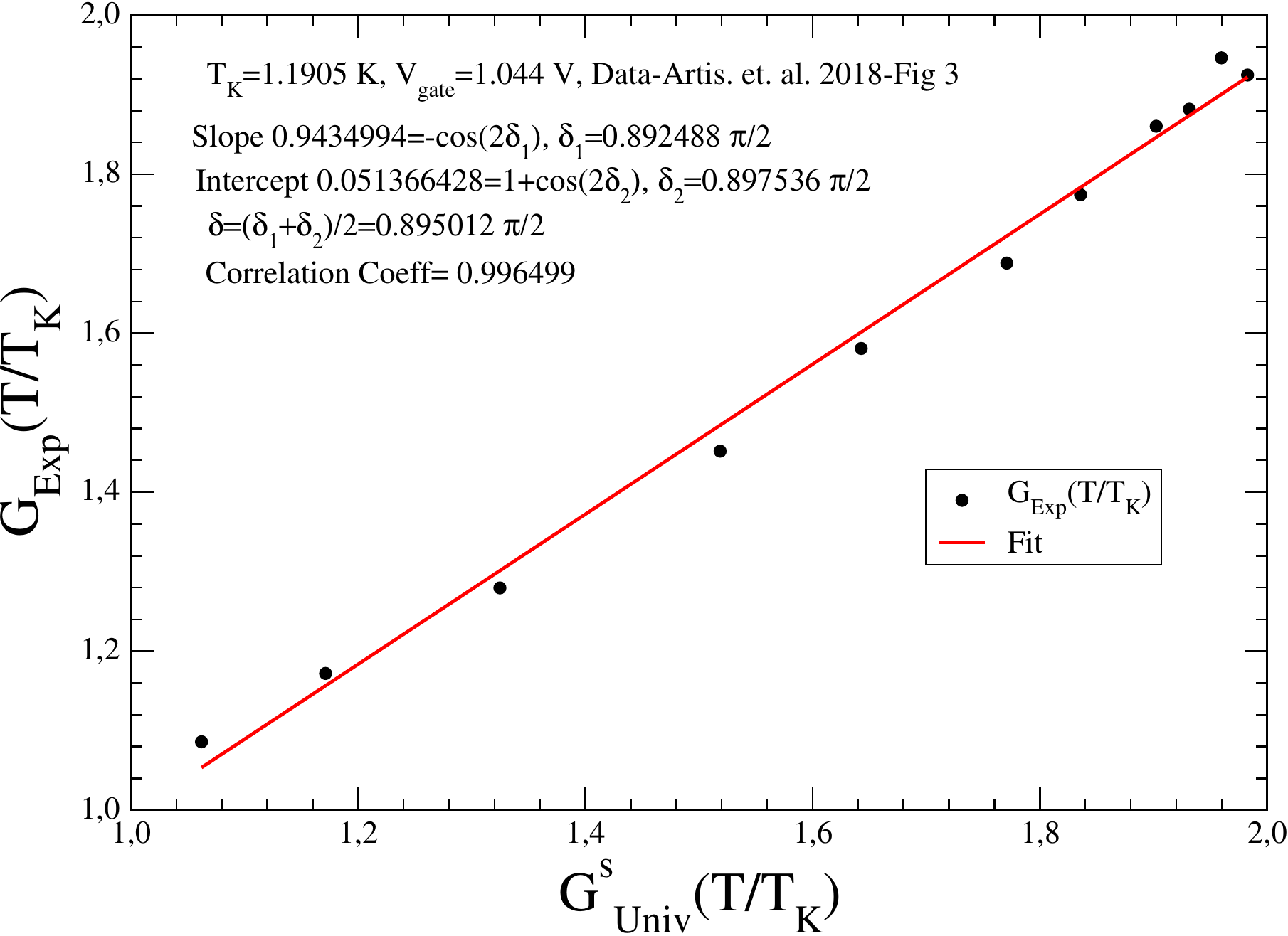}
\caption {(Color online) $G(T^{*})$ vs. $G^{S}(T^{*})$, where the experimental data were obtained from  the Artis $et$ $al.$ experimental paper \cite{Artis2018} 
with $V_{gate}=1.044V$ (see Eq. \ref{G_Seridonio}).}
\label{fig31}
\end{center}
\end{figure}

\begin{figure}[htp] 
\begin{center}
\includegraphics[clip,width=0.40\textwidth,angle=0.0]{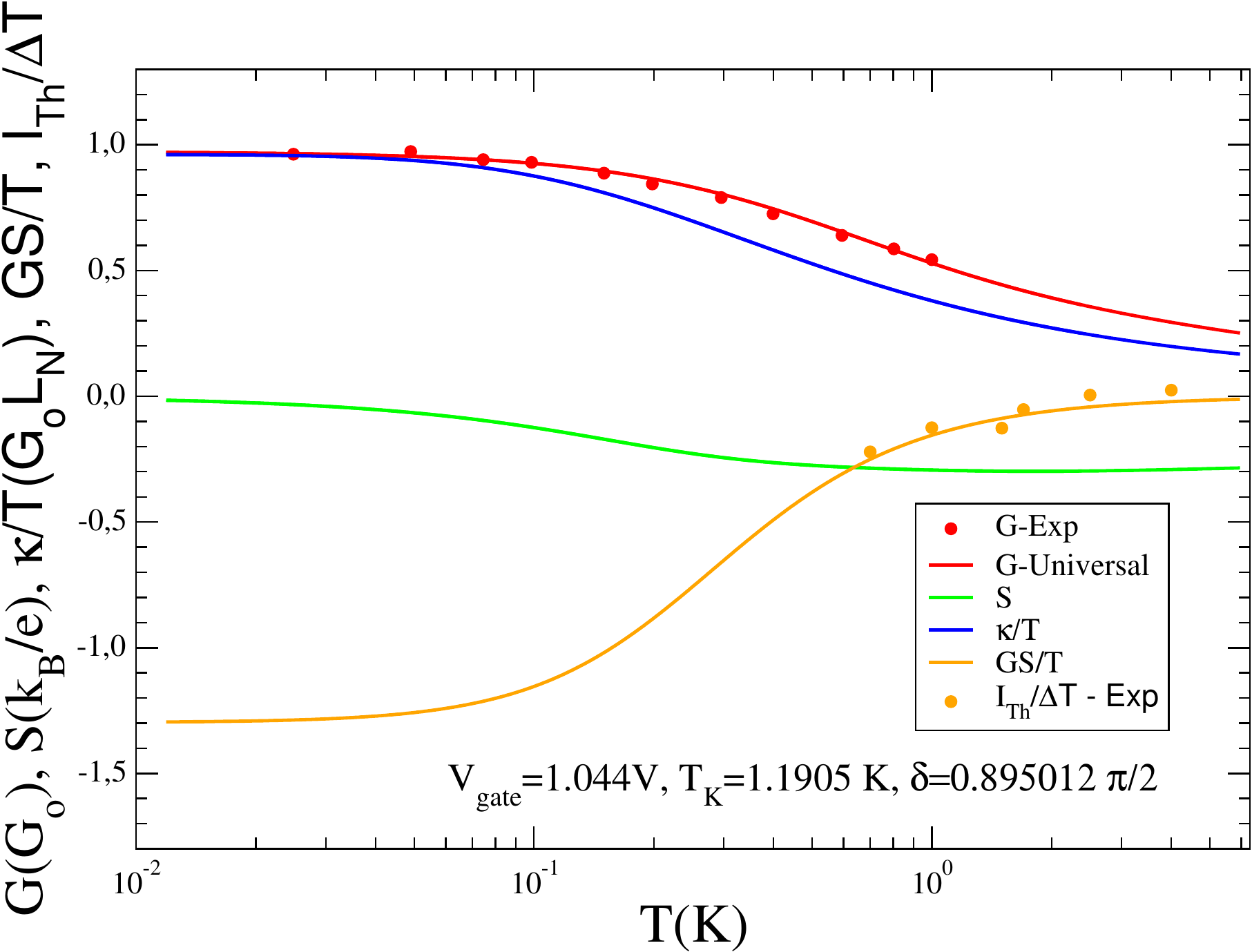}
\caption {(Color online) Thermoelectric properties as temperature function in Kelvin degrees. The experimental data were obtained from  the Artis $et$ $al.$ experimental 
paper \cite{Artis2018} with $V_{gate}=1.044V$.}
\label{fig32}
\end{center}
\end{figure}

\begin{figure}[htp] 
\begin{center}
\includegraphics[clip,width=0.40\textwidth,angle=0.0]{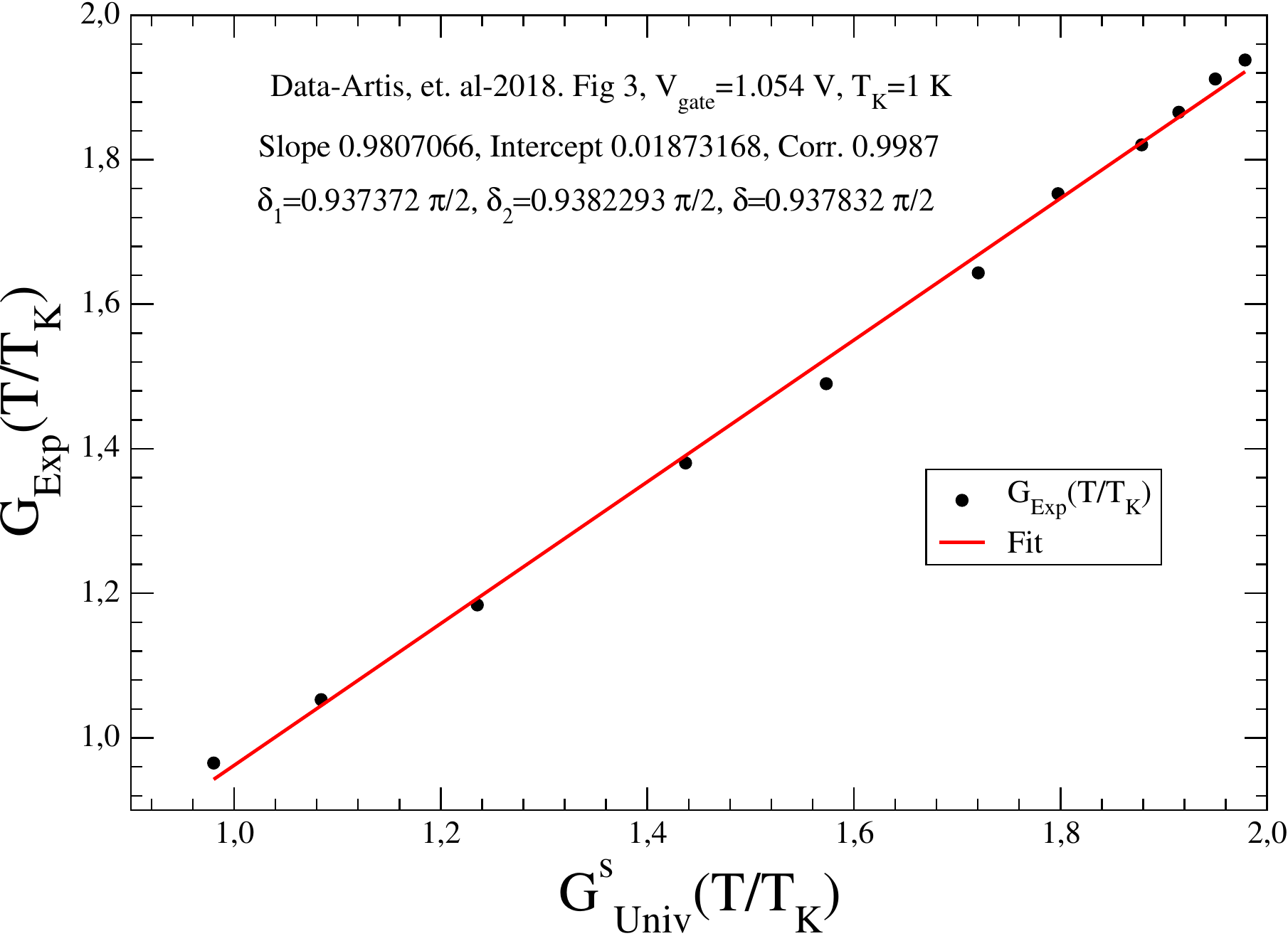}
\caption {(Color online) $G(T^{*})$ vs. $G^{S}(T^{*})$, where the experimental data were obtained from  the Artis $et$ $al.$ experimental paper \cite{Artis2018} with
$V_{gate}=1.054V$ (see Eq. \ref{G_Seridonio}).}
\label{fig33}
\end{center}
\end{figure}

\begin{figure}[htp] 
\begin{center}
\includegraphics[clip,width=0.40\textwidth,angle=0.0]{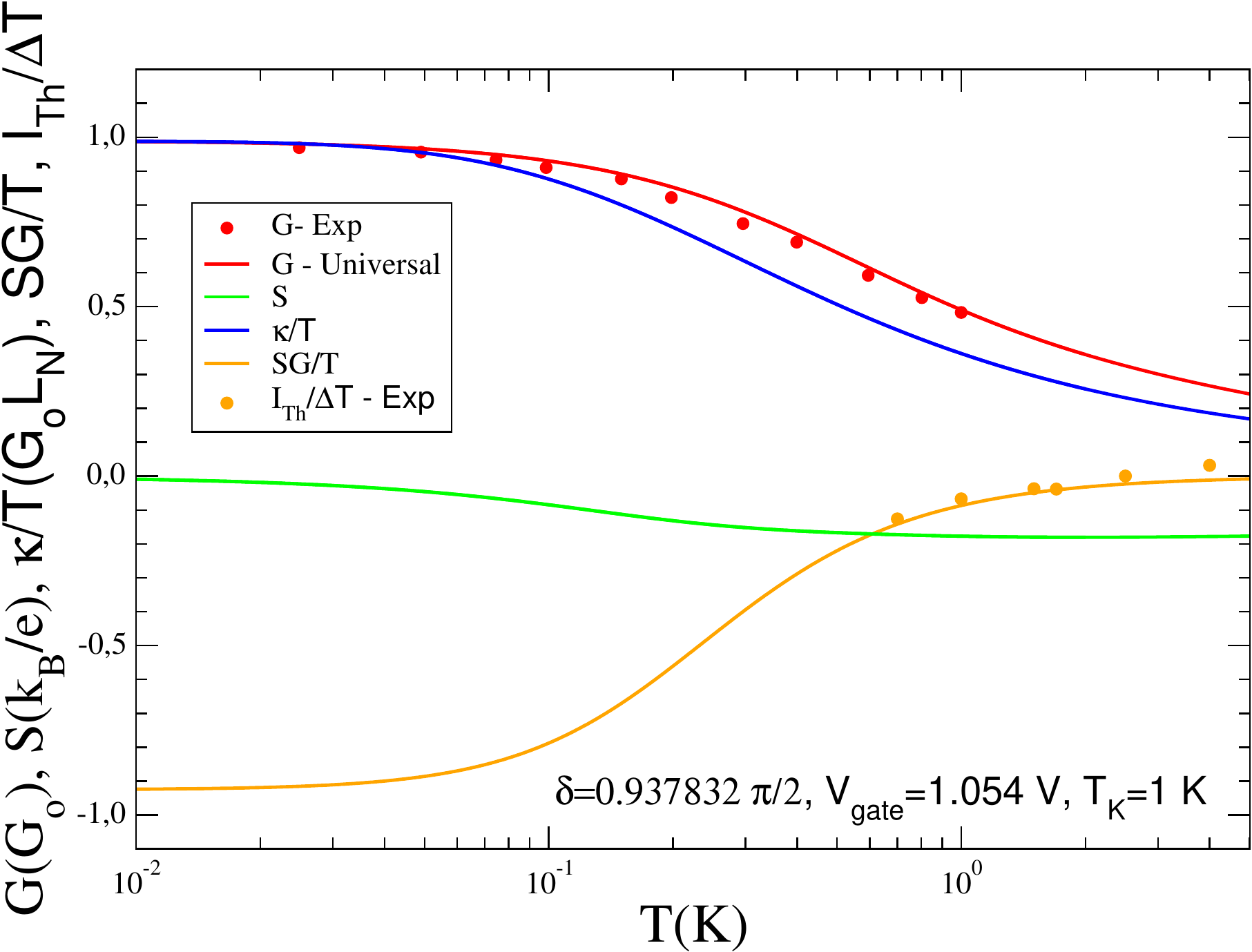}
\caption {(Color online) Thermoelectric properties as a function of temperature in Kelvin degrees. The experimental data were obtained from  the Artis $et$ $al.$ experimental 
paper \cite{Artis2018} with $V_{gate}=1.054V$.}
\label{fig34}
\end{center}
\end{figure}
Considering an ohmic dependence between the thermovoltage and 
the thermocurrent in an experimental device, the relation 
$SG=\alpha\frac{I_{Th}}{\Delta T}$ is valid, where $\Delta T$ is the difference of temperatures associated with the Seebeck effect, and where $\alpha$ must be a dimensionless 
constant for a fixed $V_{gate}$, but it could be a temperature function. If we assume that at low temperatures 
$\alpha=\eta T$, with $\eta$ being a constant that has the inverse of temperature units, we expected that 
\begin{equation}
\frac{I_{Th}}{\Delta T}\simeq \frac{SG}{T} .
\label{ITh}
\end{equation} 

To explore the validity of our predictions, we employed the following results of the Artis $et$ $al.$ paper \cite{Artis2018}: the data of figures $2b$ and $3(a,b)$ for the 
electrical conductance $G(T)$ at different gate voltages $V_{gate}$, and the figures $2c$ and $4a$
that present the results of $\frac{I_{Th}}{\Delta T}$ as $V_{gate}$ function for different temperatures. 

In Figs. \ref{fig31} and \ref{fig33}, we show the results of the slope and the intercept of the linear figures corresponding to the electrical conductance $G(T)$ of the 
Artis $et$ $al.$ \cite{Artis2018} experimental data. We obtained for $V_{gate}=1.044V$: $\delta=0.895012 \pi/2$ and $T_{K}=1.1905K$ and for 
$V_{gate}=1.054V$: $\delta=0.937832 \pi/2$ and $T_{K}=1.0K$. These Kondo temperatures agree well with the experimental results indicated in Fig. (3b) of the reference
\cite{Artis2018}. 

Employing the $\delta$ and $T_{K}$ values obtained, the universal relations for the Onsager coefficients (Eqs. \ref{Lo}-\ref{L2}) and Eqs. \ref{G}, \ref{K} and \ref{S}, 
we compute the thermal dependence of the thermoelectric properties: $G(T)$, $S(T)$, $\frac{\kappa}{T}$, and $GS/T$. 

In Fig. \ref{fig32}, we show the results of those properties, corresponding to $V_{gate}=1.044V$. Initially, we adjust the universal electrical conductance $G_{Universal}$ 
to the corresponding experimental one, $G-Exp.$, and we compute $\frac{SG}{T}$ and compare it with the experimental data of $\frac{I_{Th}}{\Delta T}$. Unfortunately, the number 
of experimental data for the $\frac{I_{Th}}{\Delta T}$ in the Artis $et$ $al.$ paper is limited, but the agreement of both properties with the available experimental data is 
excellent. Although there are no available experimental results for the temperature-normalized thermal conductance $\kappa/T$ and the thermopower $S(T)$, we calculated these 
properties and obtained fair, reliable results: the $\kappa/T$ exhibits behavior similar to Fig. \ref{fig2}, and $S(t)$ goes to zero at low temperatures, indicating that the 
experimental measurements were performed in a Kondo situation nearer the symmetric limit. 

In Fig. \ref{fig34}, we show the same results of Fig. \ref{fig32}, corresponding to $V_{gate}=1.054V$, and the results show the same overall behavior as the previous one.

\section{Conclusions}
\label{sec7}

In the present investigation, we employ the NRG treatment to compute the thermal dependence of the TTCs in the Kondo regime. From  
Eqs. \ref{Lo}-\ref{L2}, we can obtain the thermal dependence of the thermoelectric transport coefficients $L_{0}$, $L_{1}$ and $L_{2}$ in asymmetric conditions in terms of the 
Kondo temperature and the parameter $\delta$. All of the thermal dependence is ``carried" through the symmetric thermoelectric coefficients' universal functions of 
$T^{*}$, and all of the dependence on the parameters of the model is taken into account through the parameter $\delta$. We also derived simple universal fitting formulas for 
the TTCs, given by Eqs. \ref{Ajuste1}, \ref{Fit2}, and \ref{Ajuste2}, discussed in the Appendix \ref{method}, that can be used to predict the thermoelectric properties. 

In practical terms, knowledge of the experimental results of the electrical conductance or the thermopower in the Kondo regime at temperature function 
$\left\{G(T_{i}),T_{i}\right\}$ or $\left\{S(T_{i},T_{i}\right\}$, allows the determination of the Kondo temperature $T_{K}$ and the parameter $\delta$, and employing the TTCs, 
we can calculate all the other thermoelectric properties. 
The ideal situation to ``check'' our procedure is to obtain all the thermoelectric properties from the experimental measurements, but this requires a consistent and complete 
set of experimental data for $G(T)$, $S(T)$, and $\kappa(T)$, in a broad temperature range below and above the Kondo temperature, for the same $V_{gate}$. Unfortunately, 
we did not find such experimental measurements in the literature, but several papers report measured the gate dependence $V_{gate}=V$ of those properties, for a fixed temperature, 
$T$ \cite{Scheibner2005,Hoffmann2009,Svensson2013,Artis2018,Dutta2019}. 

We focused on applying the universal TTC methodology to the experimental results of the Artis $et$ $al.$ paper \cite{Artis2018}, which measured the electric conductance $G(T)$, 
thermocurrent $I_{Th}$ normalized by $\Delta T$, and the thermovoltage  $V_{th}$  as a function of the gate voltage $V$ for a fixed temperature. We adjusted the experimental 
results of the $G(T)$ and $I_{Th}/\Delta T$ employing the universal TTCs, obtaining excellent agreement. Although the Artis group did not measure the temperature-normalized 
thermal conductance $\kappa/T$ and the thermopower $S(T)$, we calculated those properties, obtaining reliable results. 

We expect that this investigation will motivate researchers to carry out experimental work in this direction, in order  to compare the procedure expounded here to 
experimental testing.

\begin{acknowledgments}
We are thankful for the financial support of the Research 
Division of the Colombia National University, Bogot\'a (DIB) and the Colombian Scientific Agency - COLCIENCIAS, the S\~ao Paulo State Research Foundation (FAPESP), the 
Brazilian National Research Council (CNPq) and Coordination of Superior Level Staff Improvement (CAPES). E. Ramos acknowledge support from COLCIENCIAS-COLFUTURO doctoral 
scholarship ``Convocatoria Doctorados Nacionales No. 617, 
2014-2''.  R. Franco is grateful for the hospitality of the IFSC-USP-S\~ao Carlos and the IF-UFF- Niter\'{o}i, where part of this work was done.
\end{acknowledgments}

\appendix

\section{Application of the universal TTCs methodology to experimental data}
\label{method}

Some earlier papers \cite{Seridonio_2009,Seridonio1_2009,Seridonio2_2009,Seridonio_2010} discussed how to employ experimental data of the thermal dependence of the electric 
conductance $G(T)$, to calculate the parameter $\delta$ to check the validity of Eq. \ref{Lo}. In particular, an almost perfect fit of $G(T)$ with experimental results was 
found in reference \cite{Munoz}. From the theoretical point of view, it is possible to adjust the experimental results by employing the TTCs in the symmetrical limit of the 
SIAM obtained from the NRG calculations. Nevertheless, for practical purposes, we can also employ the fitting formulas obtained in Eqs. \ref{Ajuste1}, \ref{Fit2} and 
\ref{Ajuste2}. 

\begin{figure}[htp] 
\begin{center}
\includegraphics[clip,width=0.45\textwidth,angle=0.0]{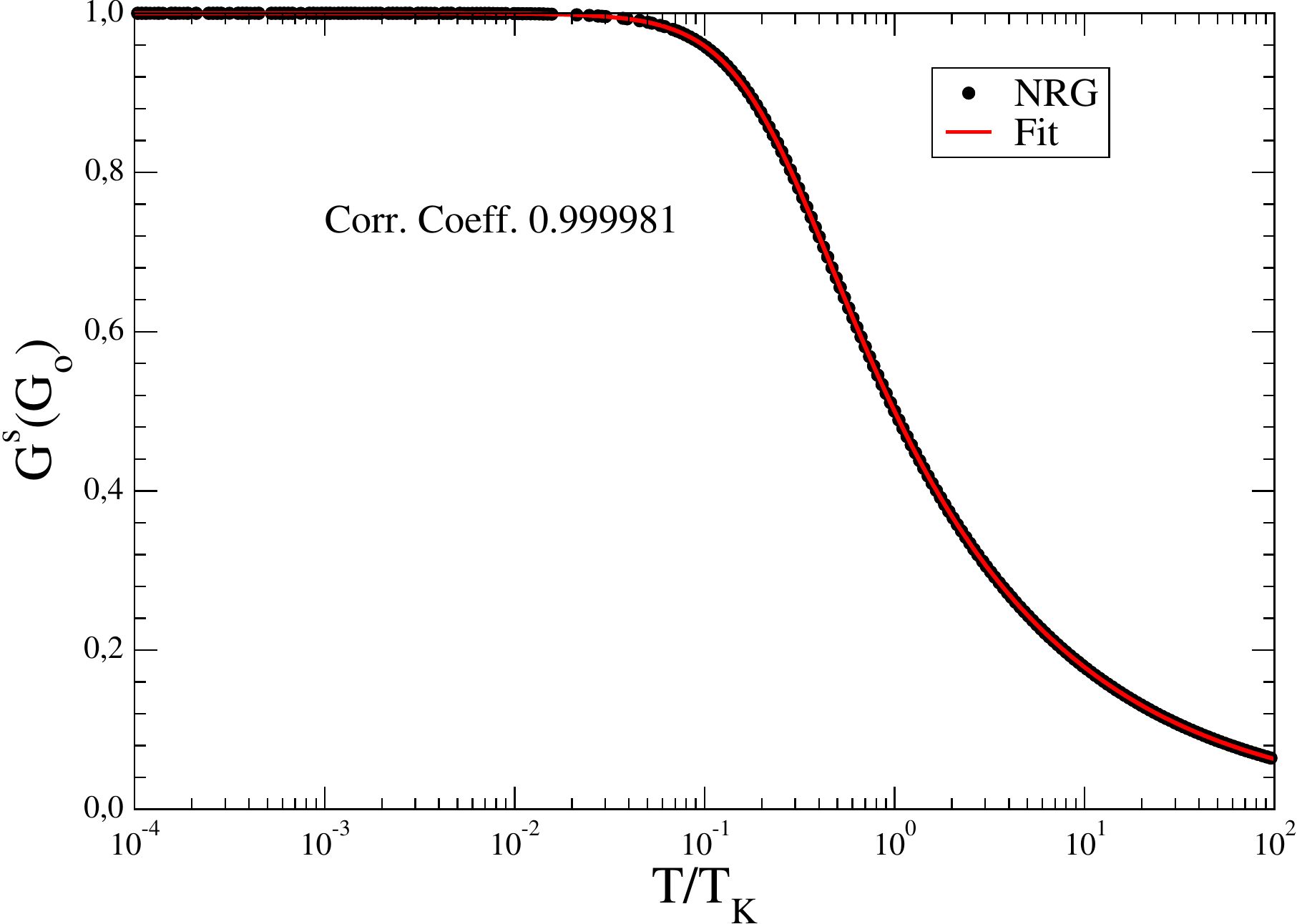}
\caption {(Color online) NRG Universal electrical conductance $G^{S}(G_{o})$ in the symmetrical limit of the SIAM vs. $\left(T^{*}\right)$, and its fit to  
Eq. \ref{Ajuste1}.}
\label{G-Uni}
\end{center}
\end{figure}

Essentially, the procedure is the following: Given a set of experimental data $\left[G_{i}\left(T_{i}\right),T_{i}\right]$, choose a trial Kondo temperature $T_{K}$, and  a 
new data set  $\left[G_{i}\left(T_{i}^{*}\right),T_{i}^{*}\right]$, with $T_{i}^{*}=\frac{T_{i}}{T_{K}}$ is generated. Since  the universal curve for the electric conductance 
in the symmetric limit of the SIAM, $G^{S}(T^{*})$ vs. $T^{*}$ is known for the fitting Eq. \ref{Ajuste1} (Fig. \ref{G-Uni}), it is possible to obtain the value 
$G^{S}_{i}(T_{i}^{*})$ for each experimental data set of $T_{i}^{*}$, and plot $G_{i}\left(T_{i}^{*}\right)$ vs. $G^{S}_{i}\left(T_{i}^{*}\right)$. If the plot followed a 
straight line, the correct Kondo temperature value $T_{K}$ was attained, and the corresponding parameter $\delta$ could be obtained by the slope and the intercept of the 
straight line (see Eq. \ref{G_Seridonio} and Figs. \ref{fig31} and \ref{fig33}). On the contrary, if the obtained plot does not follow a straight line, a new trial Kondo 
temperature $T_{K}$ must be employed, until a straight line is obtained. Employing Eq. \ref{Lo} and the fit shown in Eq. \ref{Ajuste1}, it is also possible to
compute $L_{0}(T^{*})$. 

The same procedure can be performed using the thermopower. Employing Eqs. \ref{S}, \ref{Lo} and \ref{L1}, it is possible to write the thermopower $S$ as

\begin{equation}
\frac{e}{h}S\left(T^{*}\right)=\frac{sin(2\delta)\left(\frac{L_{01}^{S}}{T}\right)\left(T^{*}\right)}{-4 cos^{2}(\delta)+2h cos(2\delta)L_{0}^{S}\left(T^{*}\right)} ,
\label{S2}
\end{equation}
which is equivalent to the equation
\begin{equation}
\frac{h}{e}\frac{\left(\frac{L_{01}^{S}}{T}\right)\left(T^{*}\right)}{S\left(T^{*}\right)}=-2cot(\delta)+2h cot(2\delta)L_{0}^{S}\left(T^{*}\right) .
\label{S3}
\end{equation}

Given a set of temperatures and a thermopower experimental data set $T_{i}$ and $S(T_{i})$ ($i=1,...,N$) ($\left[S(T_{i}),T_{i}\right]$), we can choose a tentative Kondo 
temperature $T_{K}$, compute  $T_{i}^{*}$, and obtain $\left[S_{i}(T_{i}^{*}),T_{i}^{*}\right]$. Since we know the universal functions $L_{0}^{S}(T_{i}^{*})$ 
(Fitting of $G^{S}_{0}(T_{i}^{*})$- Eq. \ref{Ajuste1} associated with Fig. \ref{G-Uni}) and $\frac{L_{01}^{S}}{T}(T_{i}^{*})$, it is then a simple matter to compute the 
fraction on the left-hand side and plot it as a function of $L_{0}^{S}(T_{i}^{*})$. If the plot is a straight line, the Kondo temperature has been found. If not, we continue 
the process until the correct value is attained. 

\begin{figure}[htp] 
\begin{center}
\includegraphics[clip,width=0.45\textwidth,angle=0.0]{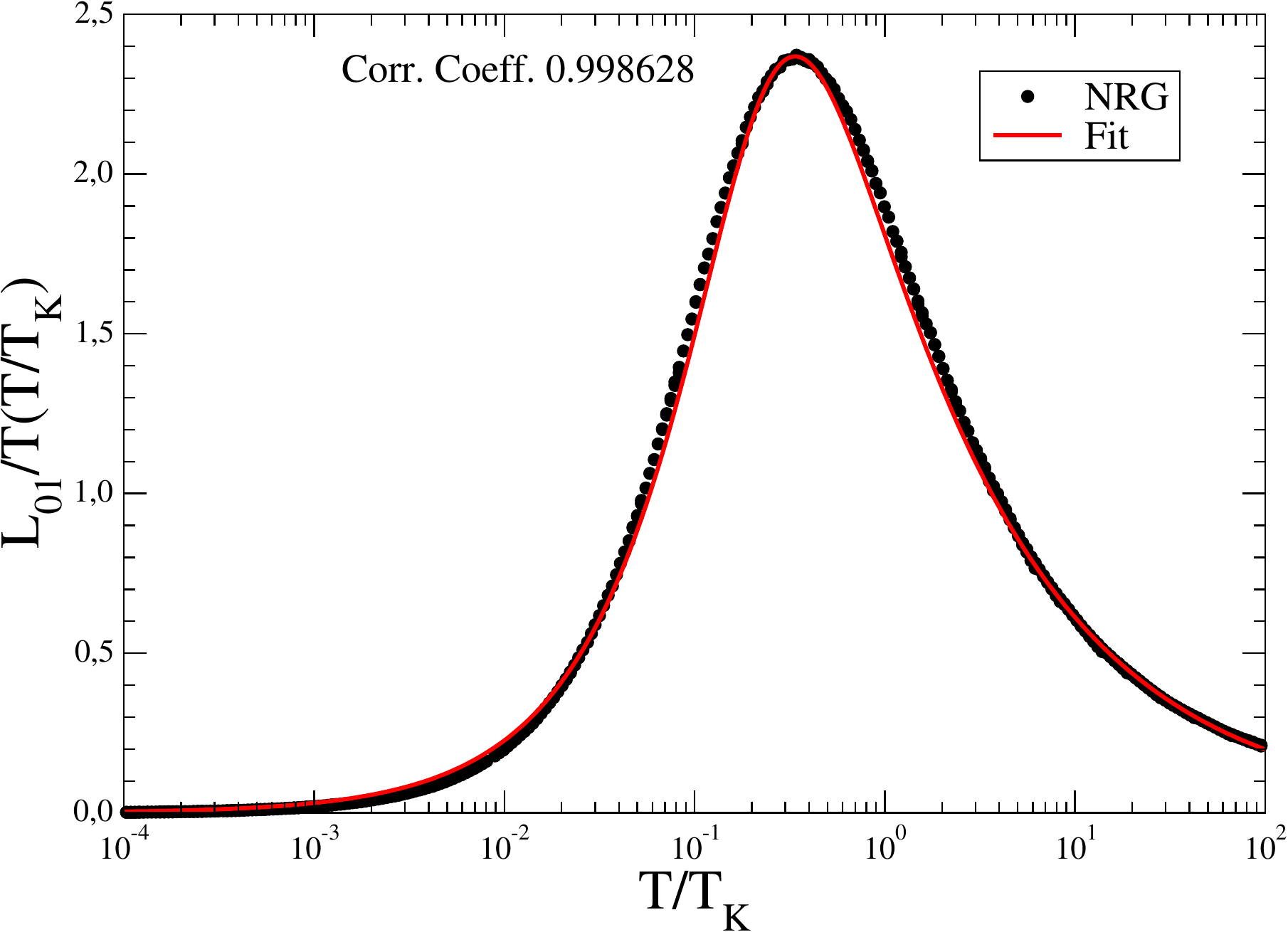}
\caption{Thermal dependence of the universal function $\left(\frac{L_{(01)}}{T}\right)$ vs. $T^{*}$, in the symmetric limit of the SIAM and its fitting to the 
Eq. \ref{Fit2}}.
\label{FitL01}
\end{center}		
\end{figure}

\begin{figure}[htp] 
\begin{center}
\includegraphics[clip,width=0.45\textwidth,angle=0.0]{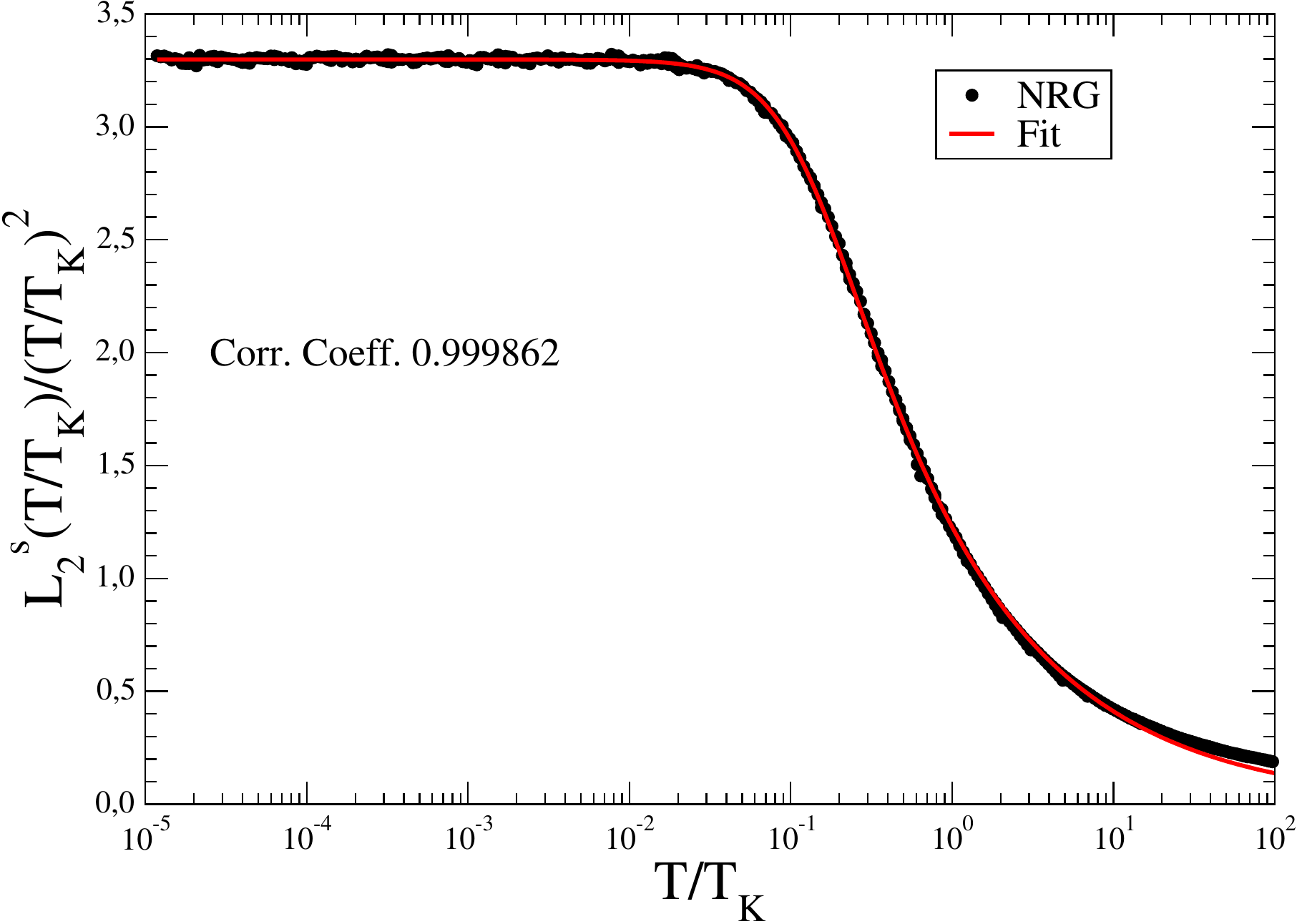}
\caption {(Color online) The universal quantity $\frac{L_{2}^{S}\left(T^{*}\right)}
{\left(\frac{k_{B}T}{T_{K}}\right)^{2}}$ vs. $T^{*}$, obtained by NRG and its fit to Eq. \ref{Ajuste2}.}
\label{L2-Uni}
\end{center}
\end{figure}

Once  the correct parameters $\delta$ and $T_{K}$ are obtained, it is possible to compute $\frac{L_{1}}{T}(T^{*})$, employing the universal function 
$\frac{L_{(10)}}{T}(T^{*})$, given by  Eq. \ref{L1} or the adjusted Eq. \ref{Fit2} of the results shown in Fig. \ref{FitL01}. Additionally, it is possible to compute 
the quantity $\frac{L_{2}\left(T^{*}\right)}{\left(k_{B}T^{*}\right)^{2}}$ by employing Eq. \ref{L2}, or the fit Eq. \ref{Ajuste2} of the results shown in Fig. 
\ref{L2-Uni}. Finally, using Eqs. \ref{G}, \ref{K}, and \ref{S}, we can calculate $G(T^{*})$, $S(T^{*})$ and $\kappa(T^{*})$, and with these, other quantities, such as $ZT$ and the 
Wiedemman-Franz law.

\begin{figure}[htp] 
\begin{center}
\includegraphics[clip,width=0.40\textwidth,angle=0.0]{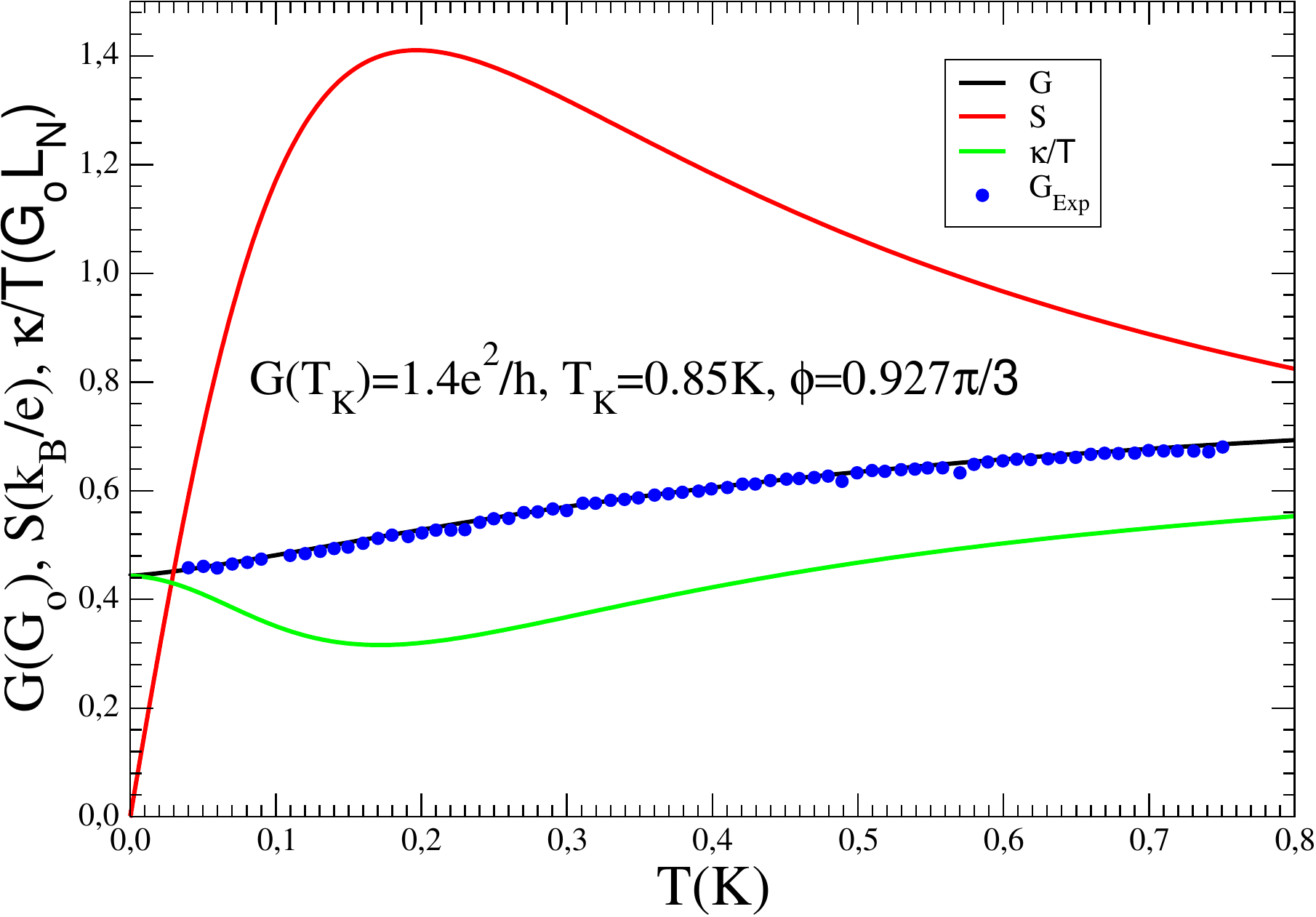}
\caption {(Color online) Electrical and thermal conductances and thermopower vs. temperature (Kelvin) for a side-coupled SET, obtained from the experimental electrical 
conductance at $T_{K}=0.85$K \cite{Sato2005}.}
\label{fig35}
\end{center}
\end{figure}

\begin{figure}[htp] 
\begin{center}
\includegraphics[clip,width=0.40\textwidth,angle=0.0]{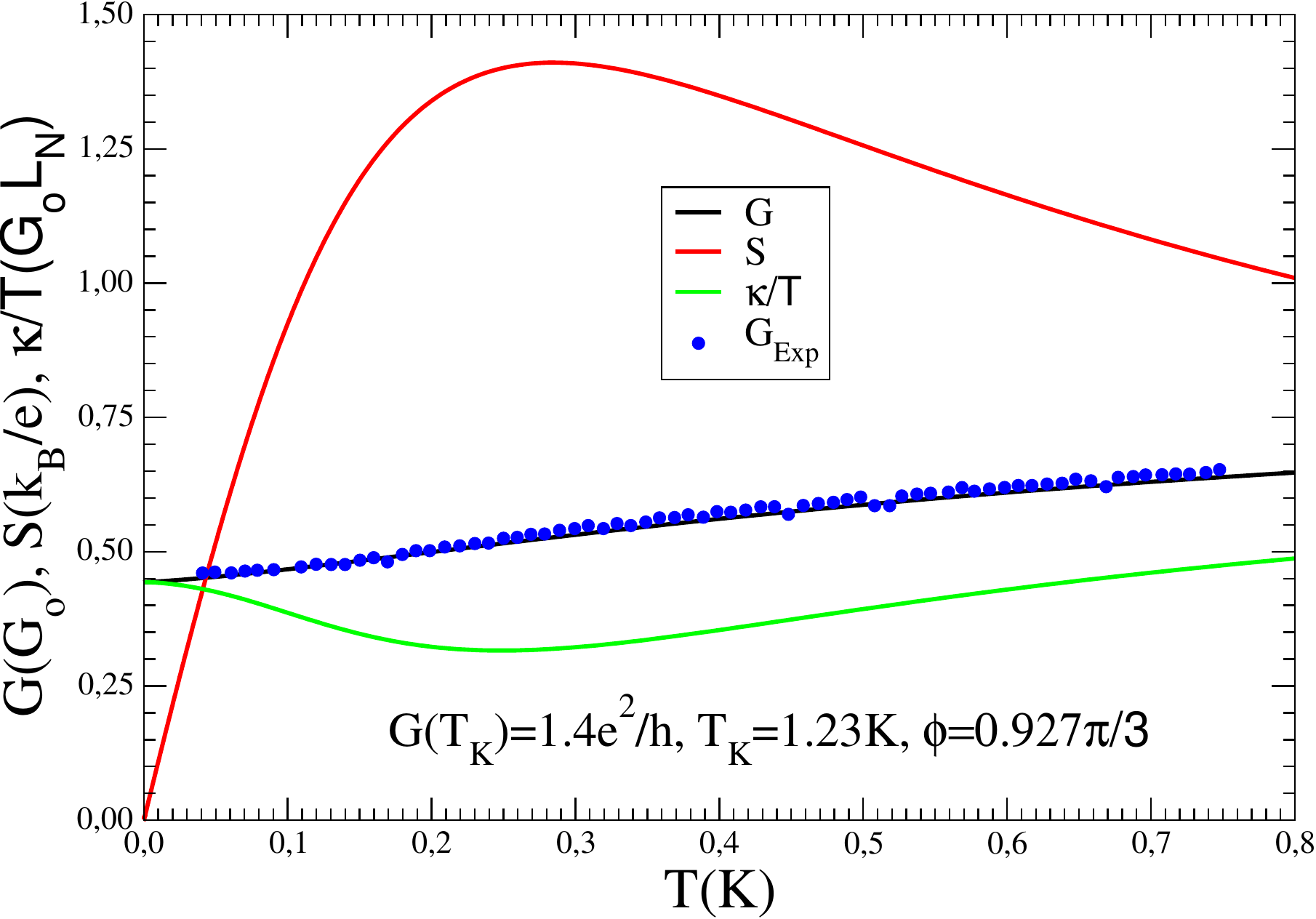}
\caption {(Color online) Electrical and thermal conductances and thermopower vs. temperature (Kelvin) for a side-coupled SET, obtained from the experimental electrical 
conductance at $T_{K}=1.23$K \cite{Sato2005}.}
\label{fig36}
\end{center}
\end{figure}   

In Fig. \ref{G-Uni}, we plot the NRG universal result \cite{Costi94} for the electrical conductance in the symmetrical limit of the Kondo regime $G^{S}(T^{*})$, employing 
the following parameters: $U=30.0D$ and $E_{f}=-15.0D$. The red line is the fit of the NRG data using a one-parameter equation employed in the reference 
\cite{Goldhaber1998PRL}, 
\begin{equation}
G(T^{*})=\frac{Go}{[(T^{*})^{2}(2^{1/\xi}-1)+1]^{\xi}} ,
\label{Ajuste1}
\end{equation}
to adjust the electrical conductance. The parameter $\xi$ determines the steepness of the decrease in conductance with increasing temperature and provides a good fit to the 
numerical renormalization group (NRG) results for the Kondo regime. In our case, $\xi=0.226022$ and the correlation coefficient$=0.999986$. The agreement achieved is excellent. 
Eq. \ref{Ajuste1}, associated with this fit, allows us to compute the universal TTC $L^{S}_{0}(T^{*})$ for any $T^{*}$ value in the range of temperatures presented.

In Fig. \ref{FitL01}, we show the thermal dependence of the universal function $\left(\frac{L_{(01)}}{T}\right)(T^{*})$ vs. $T^{*}$ in the symmetric limit of the SIAM, 
employing the following parameters: $U=30.0D$ and $E_{f}=-15.0D$. We obtain good agreement with the universal TTC $\left(\frac{L_{(01)}}{T}\right)(T^{*})$ (red line), employing
a three-parameter fit expression similar of Eq. \ref{Ajuste1} 
\begin{equation}
\left(\frac{L_{(01)}}{T}\right)\left(T^{*}\right)=A_{0}\left(\frac{(T^{*})^{\frac{A_{1}}{\xi}}}{(T^{*})^{2}
\left[A_{2}^{1/\xi}-1\right] +1}\right)^{\xi} ,
\label{Fit2}
\end{equation}
where $A_{0}=11.945007$, $A_{1}=0.860404$, $A_{2}=63.2865$ and $\xi=0.674506$.

In Fig. \ref{L2-Uni}, we plot the NRG universal results obtained for the TTC $\frac{L_{2}^{S}\left(T^{*}\right)}
{\left(k_{B}T^{*}\right)^{2}}$ in the symmetric limit of the SIAM, employing the following parameters: $U=30.0D$ and $E_{f}=-15.0D$. The red line is the fit of the 
NRG numerical data to Eq. \ref{Ajuste2}; the agreement achieved is excellent. Again, the  expression associated with this fit has a form similar to that of Eq. \ref{Ajuste1} 
\begin{equation}
\frac{L_{2}^{S}}{(T^{*})^{2}}=\frac{A_{0}}{[(T^{*})^{2}(2^{1/\xi}+A_{1})+1]^{\xi}} ,
\label{Ajuste2}
\end{equation}
where, $A_{0}=3.29776$, $\xi=0.238365$ and $A_{1}=43.6995$. This formula reduces to Eq. \ref{Ajuste1} if $A_{0}=G_{0}$ and 
$A_{1}=-1.0$. It also permits computing $\frac{L_{2}^{S}\left(T^{*}\right)}{\left(k_{B}T^{*}\right)^{2}}$ for any $T^{*}$ value in the temperature range indicated in 
the figure.  
              
For completeness, we also repeat earlier calculations employed in the paper \cite{Seridonio_2009}, which obtained the electrical conductance and the Kondo temperature 
using the experimental results of the electrical conductance SET in a side-coupled geometry \cite{Sato2005}. We show the results of those calculations in 
Figs. \ref{fig35} and \ref{fig36}. The fit of the electrical conductance is excellent. We also calculate $S(T)$ and $\kappa(T)/T$, but we cannot check the 
reliability of these results, due to the absence of experimental measurements.

\bibliography{References_thermo}

\begin{thebibliography}{45}%
\makeatletter
\providecommand \@ifxundefined [1]{%
 \@ifx{#1\undefined}
}%
\providecommand \@ifnum [1]{%
 \ifnum #1\expandafter \@firstoftwo
 \else \expandafter \@secondoftwo
 \fi
}%
\providecommand \@ifx [1]{%
 \ifx #1\expandafter \@firstoftwo
 \else \expandafter \@secondoftwo
 \fi
}%
\providecommand \natexlab [1]{#1}%
\providecommand \enquote  [1]{``#1''}%
\providecommand \bibnamefont  [1]{#1}%
\providecommand \bibfnamefont [1]{#1}%
\providecommand \citenamefont [1]{#1}%
\providecommand \href@noop [0]{\@secondoftwo}%
\providecommand \href [0]{\begingroup \@sanitize@url \@href}%
\providecommand \@href[1]{\@@startlink{#1}\@@href}%
\providecommand \@@href[1]{\endgroup#1\@@endlink}%
\providecommand \@sanitize@url [0]{\catcode `\\12\catcode `\$12\catcode
  `\&12\catcode `\#12\catcode `\^12\catcode `\_12\catcode `\%12\relax}%
\providecommand \@@startlink[1]{}%
\providecommand \@@endlink[0]{}%
\providecommand \url  [0]{\begingroup\@sanitize@url \@url }%
\providecommand \@url [1]{\endgroup\@href {#1}{\urlprefix }}%
\providecommand \urlprefix  [0]{URL }%
\providecommand \Eprint [0]{\href }%
\providecommand \doibase [0]{http://dx.doi.org/}%
\providecommand \selectlanguage [0]{\@gobble}%
\providecommand \bibinfo  [0]{\@secondoftwo}%
\providecommand \bibfield  [0]{\@secondoftwo}%
\providecommand \translation [1]{[#1]}%
\providecommand \BibitemOpen [0]{}%
\providecommand \bibitemStop [0]{}%
\providecommand \bibitemNoStop [0]{.\EOS\space}%
\providecommand \EOS [0]{\spacefactor3000\relax}%
\providecommand \BibitemShut  [1]{\csname bibitem#1\endcsname}%
\let\auto@bib@innerbib\@empty
\bibitem [{\citenamefont {S\'anchez}\ and\ \citenamefont
  {L\'opez}(2016)}]{Rosa16}%
  \BibitemOpen
  \bibfield  {author} {\bibinfo {author} {\bibfnamefont {D.}~\bibnamefont
  {S\'anchez}}\ and\ \bibinfo {author} {\bibfnamefont {R.}~\bibnamefont
  {L\'opez}},\ }\href@noop {} {\bibfield  {journal} {\bibinfo  {journal}
  {Comptes Rendus Physique}\ }\textbf {\bibinfo {volume} {17}},\ \bibinfo
  {pages} {1060 } (\bibinfo {year} {2016})}\BibitemShut {NoStop}%
\bibitem [{\citenamefont {Tritt}(2002)}]{Tritt2002}%
  \BibitemOpen
  \bibfield  {author} {\bibinfo {author} {\bibfnamefont {T.}~\bibnamefont
  {Tritt}},\ }in\ \href@noop {} {\emph {\bibinfo {booktitle} {Encyclopedia of
  Materials: Science and Technology}}},\ \bibinfo {editor} {edited by\ \bibinfo
  {editor} {\bibfnamefont {K.~J.}\ \bibnamefont {Buschow}}, \bibinfo {editor}
  {\bibfnamefont {R.~W.}\ \bibnamefont {Cahn}}, \bibinfo {editor}
  {\bibfnamefont {M.~C.}\ \bibnamefont {Flemings}}, \bibinfo {editor}
  {\bibfnamefont {B.}~\bibnamefont {Ilschner}}, \bibinfo {editor}
  {\bibfnamefont {E.~J.}\ \bibnamefont {Kramer}}, \bibinfo {editor}
  {\bibfnamefont {S.}~\bibnamefont {Mahajan}}, \ and\ \bibinfo {editor}
  {\bibfnamefont {P.}~\bibnamefont {Veyssiere}}}\ (\bibinfo  {publisher}
  {Elsevier},\ \bibinfo {address} {Oxford},\ \bibinfo {year} {2002})\
  p.~\bibinfo {pages} {1}\BibitemShut {NoStop}%
\bibitem [{\citenamefont {Joffe}\ and\ \citenamefont
  {Stil}(1959)}]{Joffe_1959}%
  \BibitemOpen
  \bibfield  {author} {\bibinfo {author} {\bibfnamefont {A.~F.}\ \bibnamefont
  {Joffe}}\ and\ \bibinfo {author} {\bibfnamefont {L.~S.}\ \bibnamefont
  {Stil}},\ }\href@noop {} {\bibfield  {journal} {\bibinfo  {journal} {Reports
  on Progress in Physics}\ }\textbf {\bibinfo {volume} {22}},\ \bibinfo {pages}
  {167} (\bibinfo {year} {1959})}\BibitemShut {NoStop}%
\bibitem [{\citenamefont {Wright}(1958)}]{Wright1958}%
  \BibitemOpen
  \bibfield  {author} {\bibinfo {author} {\bibfnamefont {D.~A.}\ \bibnamefont
  {Wright}},\ }\href {\doibase 10.1038/181834a0} {\bibfield  {journal}
  {\bibinfo  {journal} {Nature}\ }\textbf {\bibinfo {volume} {181}},\ \bibinfo
  {pages} {834} (\bibinfo {year} {1958})}\BibitemShut {NoStop}%
\bibitem [{\citenamefont {Witting}\ \emph {et~al.}(2019)\citenamefont
  {Witting}, \citenamefont {Chasapis}, \citenamefont {Ricci}, \citenamefont
  {Peters}, \citenamefont {Heinz}, \citenamefont {Hautier},\ and\ \citenamefont
  {Snyder}}]{Witting2019}%
  \BibitemOpen
  \bibfield  {author} {\bibinfo {author} {\bibfnamefont {I.~T.}\ \bibnamefont
  {Witting}}, \bibinfo {author} {\bibfnamefont {T.~C.}\ \bibnamefont
  {Chasapis}}, \bibinfo {author} {\bibfnamefont {F.}~\bibnamefont {Ricci}},
  \bibinfo {author} {\bibfnamefont {M.}~\bibnamefont {Peters}}, \bibinfo
  {author} {\bibfnamefont {N.~A.}\ \bibnamefont {Heinz}}, \bibinfo {author}
  {\bibfnamefont {G.}~\bibnamefont {Hautier}}, \ and\ \bibinfo {author}
  {\bibfnamefont {G.~J.}\ \bibnamefont {Snyder}},\ }\href@noop {} {\bibfield
  {journal} {\bibinfo  {journal} {Advanced Electronic Materials}\ }\textbf
  {\bibinfo {volume} {5}},\ \bibinfo {pages} {1800904} (\bibinfo {year}
  {2019})}\BibitemShut {NoStop}%
\bibitem [{\citenamefont {He}\ and\ \citenamefont {Tritt}(2017)}]{Jian17}%
  \BibitemOpen
  \bibfield  {author} {\bibinfo {author} {\bibfnamefont {J.}~\bibnamefont
  {He}}\ and\ \bibinfo {author} {\bibfnamefont {T.~M.}\ \bibnamefont {Tritt}},\
  }\href@noop {} {\bibfield  {journal} {\bibinfo  {journal} {Science}\ }\textbf
  {\bibinfo {volume} {357}} (\bibinfo {year} {2017})}\BibitemShut {NoStop}%
\bibitem [{\citenamefont {Benenti}\ \emph {et~al.}(2017)\citenamefont
  {Benenti}, \citenamefont {Casati}, \citenamefont {Saito},\ and\ \citenamefont
  {Whitney}}]{Benenti2017}%
  \BibitemOpen
  \bibfield  {author} {\bibinfo {author} {\bibfnamefont {G.}~\bibnamefont
  {Benenti}}, \bibinfo {author} {\bibfnamefont {G.}~\bibnamefont {Casati}},
  \bibinfo {author} {\bibfnamefont {K.}~\bibnamefont {Saito}}, \ and\ \bibinfo
  {author} {\bibfnamefont {R.}~\bibnamefont {Whitney}},\ }\href@noop {}
  {\bibfield  {journal} {\bibinfo  {journal} {Physics Reports}\ }\textbf
  {\bibinfo {volume} {694}},\ \bibinfo {pages} {1 } (\bibinfo {year}
  {2017})}\BibitemShut {NoStop}%
\bibitem [{\citenamefont {Xu}\ \emph {et~al.}(2017)\citenamefont {Xu},
  \citenamefont {Xu},\ and\ \citenamefont {Zhu}}]{Xu2017}%
  \BibitemOpen
  \bibfield  {author} {\bibinfo {author} {\bibfnamefont {N.}~\bibnamefont
  {Xu}}, \bibinfo {author} {\bibfnamefont {Y.}~\bibnamefont {Xu}}, \ and\
  \bibinfo {author} {\bibfnamefont {J.}~\bibnamefont {Zhu}},\ }\href {\doibase
  10.1038/s41535-017-0054-3} {\bibfield  {journal} {\bibinfo  {journal} {npj
  Quantum Materials}\ }\textbf {\bibinfo {volume} {2}},\ \bibinfo {pages} {51}
  (\bibinfo {year} {2017})}\BibitemShut {NoStop}%
\bibitem [{\citenamefont {Gooth}\ \emph {et~al.}(2018)\citenamefont {Gooth},
  \citenamefont {Schierning}, \citenamefont {Felser},\ and\ \citenamefont
  {Nielsch}}]{gooth2018}%
  \BibitemOpen
  \bibfield  {author} {\bibinfo {author} {\bibfnamefont {J.}~\bibnamefont
  {Gooth}}, \bibinfo {author} {\bibfnamefont {G.}~\bibnamefont {Schierning}},
  \bibinfo {author} {\bibfnamefont {C.}~\bibnamefont {Felser}}, \ and\ \bibinfo
  {author} {\bibfnamefont {K.}~\bibnamefont {Nielsch}},\ }\href {\doibase
  10.1557/mrs.2018.34} {\bibfield  {journal} {\bibinfo  {journal} {MRS
  Bulletin}\ }\textbf {\bibinfo {volume} {43}},\ \bibinfo {pages} {187}
  (\bibinfo {year} {2018})}\BibitemShut {NoStop}%
\bibitem [{\citenamefont {Zoui}\ \emph {et~al.}(2020)\citenamefont {Zoui},
  \citenamefont {S.}, \citenamefont {J.G.},\ and\ \citenamefont
  {M.}}]{Bourouis20}%
  \BibitemOpen
  \bibfield  {author} {\bibinfo {author} {\bibfnamefont {M.}~\bibnamefont
  {Zoui}}, \bibinfo {author} {\bibfnamefont {B.}~\bibnamefont {S.}}, \bibinfo
  {author} {\bibfnamefont {S.}~\bibnamefont {J.G.}}, \ and\ \bibinfo {author}
  {\bibfnamefont {B.}~\bibnamefont {M.}},\ }\href {\doibase 10.3390/en13143606}
  {\bibfield  {journal} {\bibinfo  {journal} {Energies}\ }\textbf {\bibinfo
  {volume} {13}},\ \bibinfo {pages} {3606} (\bibinfo {year}
  {2020})}\BibitemShut {NoStop}%
\bibitem [{\citenamefont {Goldhaber-Gordon}\ \emph
  {et~al.}(1998{\natexlab{a}})\citenamefont {Goldhaber-Gordon}, \citenamefont
  {Shtrikman}, \citenamefont {Mahalu}, \citenamefont {Abusch-Magder},
  \citenamefont {Meirav},\ and\ \citenamefont {Kastner}}]{Goldhaber1998}%
  \BibitemOpen
  \bibfield  {author} {\bibinfo {author} {\bibfnamefont {D.}~\bibnamefont
  {Goldhaber-Gordon}}, \bibinfo {author} {\bibfnamefont {H.}~\bibnamefont
  {Shtrikman}}, \bibinfo {author} {\bibfnamefont {D.}~\bibnamefont {Mahalu}},
  \bibinfo {author} {\bibfnamefont {D.}~\bibnamefont {Abusch-Magder}}, \bibinfo
  {author} {\bibfnamefont {U.}~\bibnamefont {Meirav}}, \ and\ \bibinfo {author}
  {\bibfnamefont {M.~A.}\ \bibnamefont {Kastner}},\ }\href {\doibase
  10.1038/34373} {\bibfield  {journal} {\bibinfo  {journal} {Nature}\ }\textbf
  {\bibinfo {volume} {391}},\ \bibinfo {pages} {156} (\bibinfo {year}
  {1998}{\natexlab{a}})}\BibitemShut {NoStop}%
\bibitem [{\citenamefont {Yoshida}\ and\ \citenamefont
  {Oliveira}(2009)}]{Yoshida2009}%
  \BibitemOpen
  \bibfield  {author} {\bibinfo {author} {\bibfnamefont {M.}~\bibnamefont
  {Yoshida}}\ and\ \bibinfo {author} {\bibfnamefont {L.}~\bibnamefont
  {Oliveira}},\ }\href {\doibase https://doi.org/10.1016/j.physb.2009.07.118}
  {\bibfield  {journal} {\bibinfo  {journal} {Physica B: Condensed Matter}\
  }\textbf {\bibinfo {volume} {404}},\ \bibinfo {pages} {3312 } (\bibinfo
  {year} {2009})}\BibitemShut {NoStop}%
\bibitem [{\citenamefont {Costi}\ and\ \citenamefont
  {Zlatic}(2010)}]{Costi2010}%
  \BibitemOpen
  \bibfield  {author} {\bibinfo {author} {\bibfnamefont {T.~A.}\ \bibnamefont
  {Costi}}\ and\ \bibinfo {author} {\bibfnamefont {V.}~\bibnamefont {Zlatic}},\
  }\href {\doibase 10.1103/PhysRevB.81.235127} {\bibfield  {journal} {\bibinfo
  {journal} {Phys. Rev. B}\ }\textbf {\bibinfo {volume} {81}},\ \bibinfo
  {pages} {235127} (\bibinfo {year} {2010})}\BibitemShut {NoStop}%
\bibitem [{\citenamefont {Hershfield}\ \emph {et~al.}(2013)\citenamefont
  {Hershfield}, \citenamefont {Muttalib},\ and\ \citenamefont
  {Nartowt}}]{Hershfield13}%
  \BibitemOpen
  \bibfield  {author} {\bibinfo {author} {\bibfnamefont {S.}~\bibnamefont
  {Hershfield}}, \bibinfo {author} {\bibfnamefont {K.~A.}\ \bibnamefont
  {Muttalib}}, \ and\ \bibinfo {author} {\bibfnamefont {B.~J.}\ \bibnamefont
  {Nartowt}},\ }\href {\doibase 10.1103/PhysRevB.88.085426} {\bibfield
  {journal} {\bibinfo  {journal} {Phys. Rev. B}\ }\textbf {\bibinfo {volume}
  {88}},\ \bibinfo {pages} {085426} (\bibinfo {year} {2013})}\BibitemShut
  {NoStop}%
\bibitem [{\citenamefont {Donsa}\ \emph {et~al.}(2014)\citenamefont {Donsa},
  \citenamefont {Andergassen},\ and\ \citenamefont {Held}}]{Donsa14}%
  \BibitemOpen
  \bibfield  {author} {\bibinfo {author} {\bibfnamefont {S.}~\bibnamefont
  {Donsa}}, \bibinfo {author} {\bibfnamefont {S.}~\bibnamefont {Andergassen}},
  \ and\ \bibinfo {author} {\bibfnamefont {K.}~\bibnamefont {Held}},\ }\href
  {\doibase 10.1103/PhysRevB.89.125103} {\bibfield  {journal} {\bibinfo
  {journal} {Phys. Rev. B}\ }\textbf {\bibinfo {volume} {89}},\ \bibinfo
  {pages} {125103} (\bibinfo {year} {2014})}\BibitemShut {NoStop}%
\bibitem [{\citenamefont {Talbo}\ \emph {et~al.}(2017)\citenamefont {Talbo},
  \citenamefont {Saint-Martin}, \citenamefont {Retailleau},\ and\ \citenamefont
  {Dollfus}}]{Talbo2017}%
  \BibitemOpen
  \bibfield  {author} {\bibinfo {author} {\bibfnamefont {V.}~\bibnamefont
  {Talbo}}, \bibinfo {author} {\bibfnamefont {J.}~\bibnamefont {Saint-Martin}},
  \bibinfo {author} {\bibfnamefont {S.}~\bibnamefont {Retailleau}}, \ and\
  \bibinfo {author} {\bibfnamefont {P.}~\bibnamefont {Dollfus}},\ }\href
  {\doibase 10.1038/s41598-017-14009-4} {\bibfield  {journal} {\bibinfo
  {journal} {Scientific Reports}\ }\textbf {\bibinfo {volume} {7}},\ \bibinfo
  {pages} {14783} (\bibinfo {year} {2017})}\BibitemShut {NoStop}%
\bibitem [{\citenamefont {Costi}(2019{\natexlab{a}})}]{Costi20191}%
  \BibitemOpen
  \bibfield  {author} {\bibinfo {author} {\bibfnamefont {T.~A.}\ \bibnamefont
  {Costi}},\ }\href {\doibase 10.1103/PhysRevB.100.161106} {\bibfield
  {journal} {\bibinfo  {journal} {Phys. Rev. B}\ }\textbf {\bibinfo {volume}
  {100}},\ \bibinfo {pages} {161106} (\bibinfo {year}
  {2019}{\natexlab{a}})}\BibitemShut {NoStop}%
\bibitem [{\citenamefont {Costi}(2019{\natexlab{b}})}]{Costi20192}%
  \BibitemOpen
  \bibfield  {author} {\bibinfo {author} {\bibfnamefont {T.~A.}\ \bibnamefont
  {Costi}},\ }\href {\doibase 10.1103/PhysRevB.100.155126} {\bibfield
  {journal} {\bibinfo  {journal} {Phys. Rev. B}\ }\textbf {\bibinfo {volume}
  {100}},\ \bibinfo {pages} {155126} (\bibinfo {year}
  {2019}{\natexlab{b}})}\BibitemShut {NoStop}%
\bibitem [{\citenamefont {Kleeorin}\ \emph {et~al.}(2019)\citenamefont
  {Kleeorin}, \citenamefont {Thierschmann}, \citenamefont {Buhmann},
  \citenamefont {Georges}, \citenamefont {Molenkamp},\ and\ \citenamefont
  {Meir}}]{Thierschmann2019}%
  \BibitemOpen
  \bibfield  {author} {\bibinfo {author} {\bibfnamefont {Y.}~\bibnamefont
  {Kleeorin}}, \bibinfo {author} {\bibfnamefont {H.}~\bibnamefont
  {Thierschmann}}, \bibinfo {author} {\bibfnamefont {H.}~\bibnamefont
  {Buhmann}}, \bibinfo {author} {\bibfnamefont {A.}~\bibnamefont {Georges}},
  \bibinfo {author} {\bibfnamefont {L.~W.}\ \bibnamefont {Molenkamp}}, \ and\
  \bibinfo {author} {\bibfnamefont {Y.}~\bibnamefont {Meir}},\ }\href@noop {}
  {\bibfield  {journal} {\bibinfo  {journal} {Nature Communications}\ }\textbf
  {\bibinfo {volume} {10}},\ \bibinfo {pages} {5801} (\bibinfo {year}
  {2019})}\BibitemShut {NoStop}%
\bibitem [{\citenamefont {Eckern}\ and\ \citenamefont
  {Wysoki{\'{n}}ski}(2020)}]{Eckern2020}%
  \BibitemOpen
  \bibfield  {author} {\bibinfo {author} {\bibfnamefont {U.}~\bibnamefont
  {Eckern}}\ and\ \bibinfo {author} {\bibfnamefont {K.~I.}\ \bibnamefont
  {Wysoki{\'{n}}ski}},\ }\href {\doibase 10.1088/1367-2630/ab6874} {\bibfield
  {journal} {\bibinfo  {journal} {New Journal of Physics}\ }\textbf {\bibinfo
  {volume} {22}},\ \bibinfo {pages} {013045} (\bibinfo {year}
  {2020})}\BibitemShut {NoStop}%
\bibitem [{\citenamefont {Heremans}\ \emph {et~al.}(2004)\citenamefont
  {Heremans}, \citenamefont {Thrush},\ and\ \citenamefont
  {Morelli}}]{Heremans2004}%
  \BibitemOpen
  \bibfield  {author} {\bibinfo {author} {\bibfnamefont {J.~P.}\ \bibnamefont
  {Heremans}}, \bibinfo {author} {\bibfnamefont {C.~M.}\ \bibnamefont
  {Thrush}}, \ and\ \bibinfo {author} {\bibfnamefont {D.~T.}\ \bibnamefont
  {Morelli}},\ }\href {\doibase 10.1103/PhysRevB.70.115334} {\bibfield
  {journal} {\bibinfo  {journal} {Phys. Rev. B}\ }\textbf {\bibinfo {volume}
  {70}},\ \bibinfo {pages} {115334} (\bibinfo {year} {2004})}\BibitemShut
  {NoStop}%
\bibitem [{\citenamefont {Scheibner}\ \emph {et~al.}(2007)\citenamefont
  {Scheibner}, \citenamefont {Novik}, \citenamefont {Borzenko}, \citenamefont
  {K\"onig}, \citenamefont {Reuter}, \citenamefont {Wieck}, \citenamefont
  {Buhmann},\ and\ \citenamefont {Molenkamp}}]{Scheibner2007}%
  \BibitemOpen
  \bibfield  {author} {\bibinfo {author} {\bibfnamefont {R.}~\bibnamefont
  {Scheibner}}, \bibinfo {author} {\bibfnamefont {E.~G.}\ \bibnamefont
  {Novik}}, \bibinfo {author} {\bibfnamefont {T.}~\bibnamefont {Borzenko}},
  \bibinfo {author} {\bibfnamefont {M.}~\bibnamefont {K\"onig}}, \bibinfo
  {author} {\bibfnamefont {D.}~\bibnamefont {Reuter}}, \bibinfo {author}
  {\bibfnamefont {A.~D.}\ \bibnamefont {Wieck}}, \bibinfo {author}
  {\bibfnamefont {H.}~\bibnamefont {Buhmann}}, \ and\ \bibinfo {author}
  {\bibfnamefont {L.~W.}\ \bibnamefont {Molenkamp}},\ }\href {\doibase
  10.1103/PhysRevB.75.041301} {\bibfield  {journal} {\bibinfo  {journal} {Phys.
  Rev. B}\ }\textbf {\bibinfo {volume} {75}},\ \bibinfo {pages} {041301}
  (\bibinfo {year} {2007})}\BibitemShut {NoStop}%
\bibitem [{\citenamefont {Hoffmann}\ \emph {et~al.}(2009)\citenamefont
  {Hoffmann}, \citenamefont {Nilsson}, \citenamefont {Matthews}, \citenamefont
  {Nakpathomkun}, \citenamefont {Persson}, \citenamefont {Samuelson},\ and\
  \citenamefont {Linke}}]{Hoffmann2009}%
  \BibitemOpen
  \bibfield  {author} {\bibinfo {author} {\bibfnamefont {E.~A.}\ \bibnamefont
  {Hoffmann}}, \bibinfo {author} {\bibfnamefont {H.~A.}\ \bibnamefont
  {Nilsson}}, \bibinfo {author} {\bibfnamefont {J.~E.}\ \bibnamefont
  {Matthews}}, \bibinfo {author} {\bibfnamefont {N.}~\bibnamefont
  {Nakpathomkun}}, \bibinfo {author} {\bibfnamefont {A.~I.}\ \bibnamefont
  {Persson}}, \bibinfo {author} {\bibfnamefont {L.}~\bibnamefont {Samuelson}},
  \ and\ \bibinfo {author} {\bibfnamefont {H.}~\bibnamefont {Linke}},\
  }\href@noop {} {\bibfield  {journal} {\bibinfo  {journal} {Nano Letters}\
  }\textbf {\bibinfo {volume} {9}},\ \bibinfo {pages} {779} (\bibinfo {year}
  {2009})}\BibitemShut {NoStop}%
\bibitem [{\citenamefont {Hartman}\ \emph {et~al.}(2018)\citenamefont
  {Hartman}, \citenamefont {Olsen}, \citenamefont {L{\"u}scher}, \citenamefont
  {Samani}, \citenamefont {Fallahi}, \citenamefont {Gardner}, \citenamefont
  {Manfra},\ and\ \citenamefont {Folk}}]{Hartman2018}%
  \BibitemOpen
  \bibfield  {author} {\bibinfo {author} {\bibfnamefont {N.}~\bibnamefont
  {Hartman}}, \bibinfo {author} {\bibfnamefont {C.}~\bibnamefont {Olsen}},
  \bibinfo {author} {\bibfnamefont {S.}~\bibnamefont {L{\"u}scher}}, \bibinfo
  {author} {\bibfnamefont {M.}~\bibnamefont {Samani}}, \bibinfo {author}
  {\bibfnamefont {S.}~\bibnamefont {Fallahi}}, \bibinfo {author} {\bibfnamefont
  {G.~C.}\ \bibnamefont {Gardner}}, \bibinfo {author} {\bibfnamefont
  {M.}~\bibnamefont {Manfra}}, \ and\ \bibinfo {author} {\bibfnamefont
  {J.}~\bibnamefont {Folk}},\ }\href {\doibase 10.1038/s41567-018-0250-5}
  {\bibfield  {journal} {\bibinfo  {journal} {Nature Physics}\ }\textbf
  {\bibinfo {volume} {14}},\ \bibinfo {pages} {1083} (\bibinfo {year}
  {2018})}\BibitemShut {NoStop}%
\bibitem [{\citenamefont {Svilans}\ \emph {et~al.}(2018)\citenamefont
  {Svilans}, \citenamefont {Josefsson}, \citenamefont {Burke}, \citenamefont
  {Fahlvik}, \citenamefont {Thelander}, \citenamefont {Linke},\ and\
  \citenamefont {Leijnse}}]{Artis2018}%
  \BibitemOpen
  \bibfield  {author} {\bibinfo {author} {\bibfnamefont {A.}~\bibnamefont
  {Svilans}}, \bibinfo {author} {\bibfnamefont {M.}~\bibnamefont {Josefsson}},
  \bibinfo {author} {\bibfnamefont {A.~M.}\ \bibnamefont {Burke}}, \bibinfo
  {author} {\bibfnamefont {S.}~\bibnamefont {Fahlvik}}, \bibinfo {author}
  {\bibfnamefont {C.}~\bibnamefont {Thelander}}, \bibinfo {author}
  {\bibfnamefont {H.}~\bibnamefont {Linke}}, \ and\ \bibinfo {author}
  {\bibfnamefont {M.}~\bibnamefont {Leijnse}},\ }\href {\doibase
  10.1103/PhysRevLett.121.206801} {\bibfield  {journal} {\bibinfo  {journal}
  {Phys. Rev. Lett.}\ }\textbf {\bibinfo {volume} {121}},\ \bibinfo {pages}
  {206801} (\bibinfo {year} {2018})}\BibitemShut {NoStop}%
\bibitem [{\citenamefont {Dutta}\ \emph {et~al.}(2017)\citenamefont {Dutta},
  \citenamefont {Peltonen}, \citenamefont {Antonenko}, \citenamefont {Meschke},
  \citenamefont {Skvortsov}, \citenamefont {Kubala}, \citenamefont {K\"onig},
  \citenamefont {Winkelmann}, \citenamefont {Courtois},\ and\ \citenamefont
  {Pekola}}]{Dutta2017}%
  \BibitemOpen
  \bibfield  {author} {\bibinfo {author} {\bibfnamefont {B.}~\bibnamefont
  {Dutta}}, \bibinfo {author} {\bibfnamefont {J.~T.}\ \bibnamefont {Peltonen}},
  \bibinfo {author} {\bibfnamefont {D.~S.}\ \bibnamefont {Antonenko}}, \bibinfo
  {author} {\bibfnamefont {M.}~\bibnamefont {Meschke}}, \bibinfo {author}
  {\bibfnamefont {M.~A.}\ \bibnamefont {Skvortsov}}, \bibinfo {author}
  {\bibfnamefont {B.}~\bibnamefont {Kubala}}, \bibinfo {author} {\bibfnamefont
  {J.}~\bibnamefont {K\"onig}}, \bibinfo {author} {\bibfnamefont {C.~B.}\
  \bibnamefont {Winkelmann}}, \bibinfo {author} {\bibfnamefont
  {H.}~\bibnamefont {Courtois}}, \ and\ \bibinfo {author} {\bibfnamefont
  {J.~P.}\ \bibnamefont {Pekola}},\ }\href {\doibase
  10.1103/PhysRevLett.119.077701} {\bibfield  {journal} {\bibinfo  {journal}
  {Phys. Rev. Lett.}\ }\textbf {\bibinfo {volume} {119}},\ \bibinfo {pages}
  {077701} (\bibinfo {year} {2017})}\BibitemShut {NoStop}%
\bibitem [{\citenamefont {Dutta}\ \emph {et~al.}(2019)\citenamefont {Dutta},
  \citenamefont {Majidi}, \citenamefont {Garc{\'i}a~Corral}, \citenamefont
  {Erdman}, \citenamefont {Florens}, \citenamefont {Costi}, \citenamefont
  {Courtois},\ and\ \citenamefont {Winkelmann}}]{Dutta2019}%
  \BibitemOpen
  \bibfield  {author} {\bibinfo {author} {\bibfnamefont {B.}~\bibnamefont
  {Dutta}}, \bibinfo {author} {\bibfnamefont {D.}~\bibnamefont {Majidi}},
  \bibinfo {author} {\bibfnamefont {A.}~\bibnamefont {Garc{\'i}a~Corral}},
  \bibinfo {author} {\bibfnamefont {P.~A.}\ \bibnamefont {Erdman}}, \bibinfo
  {author} {\bibfnamefont {S.}~\bibnamefont {Florens}}, \bibinfo {author}
  {\bibfnamefont {T.~A.}\ \bibnamefont {Costi}}, \bibinfo {author}
  {\bibfnamefont {H.}~\bibnamefont {Courtois}}, \ and\ \bibinfo {author}
  {\bibfnamefont {C.~B.}\ \bibnamefont {Winkelmann}},\ }\href {\doibase
  10.1021/acs.nanolett.8b04398} {\bibfield  {journal} {\bibinfo  {journal}
  {Nano Letters}\ }\textbf {\bibinfo {volume} {19}},\ \bibinfo {pages} {506}
  (\bibinfo {year} {2019})}\BibitemShut {NoStop}%
\bibitem [{\citenamefont {Hewson}(1993)}]{Hewson}%
  \BibitemOpen
  \bibfield  {author} {\bibinfo {author} {\bibfnamefont {A.~C.}\ \bibnamefont
  {Hewson}},\ }\href@noop {} {\bibfield  {journal} {\bibinfo  {journal} {The
  Kondo problem to Heavy Fermions - Cambridge University Press}\ } (\bibinfo
  {year} {1993})}\BibitemShut {NoStop}%
\bibitem [{\citenamefont {Costi}\ \emph {et~al.}(1994)\citenamefont {Costi},
  \citenamefont {Hewson},\ and\ \citenamefont {Zlatic}}]{Costi94}%
  \BibitemOpen
  \bibfield  {author} {\bibinfo {author} {\bibfnamefont {T.~A.}\ \bibnamefont
  {Costi}}, \bibinfo {author} {\bibfnamefont {A.~C.}\ \bibnamefont {Hewson}}, \
  and\ \bibinfo {author} {\bibfnamefont {V.}~\bibnamefont {Zlatic}},\ }\href
  {\doibase 10.1088/0953-8984/6/13/013} {\bibfield  {journal} {\bibinfo
  {journal} {Journal of Physics: Condensed Matter}\ }\textbf {\bibinfo {volume}
  {6}},\ \bibinfo {pages} {2519} (\bibinfo {year} {1994})}\BibitemShut
  {NoStop}%
\bibitem [{\citenamefont {Yoshida}\ \emph {et~al.}(2009)\citenamefont
  {Yoshida}, \citenamefont {Seridonio},\ and\ \citenamefont
  {Oliveira}}]{Seridonio1_2009}%
  \BibitemOpen
  \bibfield  {author} {\bibinfo {author} {\bibfnamefont {M.}~\bibnamefont
  {Yoshida}}, \bibinfo {author} {\bibfnamefont {A.~C.}\ \bibnamefont
  {Seridonio}}, \ and\ \bibinfo {author} {\bibfnamefont {L.~N.}\ \bibnamefont
  {Oliveira}},\ }\href {\doibase 10.1103/PhysRevB.80.235317} {\bibfield
  {journal} {\bibinfo  {journal} {Phys. Rev. B}\ }\textbf {\bibinfo {volume}
  {80}},\ \bibinfo {pages} {235317} (\bibinfo {year} {2009})}\BibitemShut
  {NoStop}%
\bibitem [{\citenamefont {Seridonio}\ \emph
  {et~al.}(2009{\natexlab{a}})\citenamefont {Seridonio}, \citenamefont
  {Yoshida},\ and\ \citenamefont {Oliveira}}]{Seridonio2_2009}%
  \BibitemOpen
  \bibfield  {author} {\bibinfo {author} {\bibfnamefont {A.~C.}\ \bibnamefont
  {Seridonio}}, \bibinfo {author} {\bibfnamefont {M.}~\bibnamefont {Yoshida}},
  \ and\ \bibinfo {author} {\bibfnamefont {L.~N.}\ \bibnamefont {Oliveira}},\
  }\href {\doibase 10.1103/PhysRevB.80.235318} {\bibfield  {journal} {\bibinfo
  {journal} {Phys. Rev. B}\ }\textbf {\bibinfo {volume} {80}},\ \bibinfo
  {pages} {235318} (\bibinfo {year} {2009}{\natexlab{a}})}\BibitemShut
  {NoStop}%
\bibitem [{\citenamefont {Langreth}(1966)}]{Langreth66}%
  \BibitemOpen
  \bibfield  {author} {\bibinfo {author} {\bibfnamefont {D.~C.}\ \bibnamefont
  {Langreth}},\ }\href {\doibase 10.1103/PhysRev.150.516} {\bibfield  {journal}
  {\bibinfo  {journal} {Phys. Rev.}\ }\textbf {\bibinfo {volume} {150}},\
  \bibinfo {pages} {516} (\bibinfo {year} {1966})}\BibitemShut {NoStop}%
\bibitem [{\citenamefont {Grobis}\ \emph {et~al.}(2008)\citenamefont {Grobis},
  \citenamefont {Rau}, \citenamefont {Potok}, \citenamefont {Shtrikman},\ and\
  \citenamefont {Goldhaber-Gordon}}]{Grobis}%
  \BibitemOpen
  \bibfield  {author} {\bibinfo {author} {\bibfnamefont {M.}~\bibnamefont
  {Grobis}}, \bibinfo {author} {\bibfnamefont {I.~G.}\ \bibnamefont {Rau}},
  \bibinfo {author} {\bibfnamefont {R.~M.}\ \bibnamefont {Potok}}, \bibinfo
  {author} {\bibfnamefont {H.}~\bibnamefont {Shtrikman}}, \ and\ \bibinfo
  {author} {\bibfnamefont {D.}~\bibnamefont {Goldhaber-Gordon}},\ }\href
  {\doibase 10.1103/PhysRevLett.100.246601} {\bibfield  {journal} {\bibinfo
  {journal} {Phys. Rev. Lett.}\ }\textbf {\bibinfo {volume} {100}},\ \bibinfo
  {pages} {246601} (\bibinfo {year} {2008})}\BibitemShut {NoStop}%
\bibitem [{\citenamefont {Kretinin}\ \emph {et~al.}(2011)\citenamefont
  {Kretinin}, \citenamefont {Shtrikman}, \citenamefont {Goldhaber-Gordon},
  \citenamefont {Hanl}, \citenamefont {Weichselbaum}, \citenamefont {von
  Delft}, \citenamefont {Costi},\ and\ \citenamefont {Mahalu}}]{Kretinin}%
  \BibitemOpen
  \bibfield  {author} {\bibinfo {author} {\bibfnamefont {A.~V.}\ \bibnamefont
  {Kretinin}}, \bibinfo {author} {\bibfnamefont {H.}~\bibnamefont {Shtrikman}},
  \bibinfo {author} {\bibfnamefont {D.}~\bibnamefont {Goldhaber-Gordon}},
  \bibinfo {author} {\bibfnamefont {M.}~\bibnamefont {Hanl}}, \bibinfo {author}
  {\bibfnamefont {A.}~\bibnamefont {Weichselbaum}}, \bibinfo {author}
  {\bibfnamefont {J.}~\bibnamefont {von Delft}}, \bibinfo {author}
  {\bibfnamefont {T.}~\bibnamefont {Costi}}, \ and\ \bibinfo {author}
  {\bibfnamefont {D.}~\bibnamefont {Mahalu}},\ }\href@noop {} {\bibfield
  {journal} {\bibinfo  {journal} {Phys. Rev. B}\ }\textbf {\bibinfo {volume}
  {84}},\ \bibinfo {pages} {245316} (\bibinfo {year} {2011})}\BibitemShut
  {NoStop}%
\bibitem [{\citenamefont {Parks}\ \emph {et~al.}(2010)\citenamefont {Parks},
  \citenamefont {Champagne}, \citenamefont {Costi}, \citenamefont {Shum},
  \citenamefont {Pasupathy}, \citenamefont {Neuscamman}, \citenamefont
  {Flores-Torres}, \citenamefont {Cornaglia}, \citenamefont {Aligia},
  \citenamefont {Balseiro}, \citenamefont {Chan}, \citenamefont {Abru{\~n}a},\
  and\ \citenamefont {Ralph}}]{Park}%
  \BibitemOpen
  \bibfield  {author} {\bibinfo {author} {\bibfnamefont {J.~J.}\ \bibnamefont
  {Parks}}, \bibinfo {author} {\bibfnamefont {A.~R.}\ \bibnamefont
  {Champagne}}, \bibinfo {author} {\bibfnamefont {T.~A.}\ \bibnamefont
  {Costi}}, \bibinfo {author} {\bibfnamefont {W.~W.}\ \bibnamefont {Shum}},
  \bibinfo {author} {\bibfnamefont {A.~N.}\ \bibnamefont {Pasupathy}}, \bibinfo
  {author} {\bibfnamefont {E.}~\bibnamefont {Neuscamman}}, \bibinfo {author}
  {\bibfnamefont {S.}~\bibnamefont {Flores-Torres}}, \bibinfo {author}
  {\bibfnamefont {P.~S.}\ \bibnamefont {Cornaglia}}, \bibinfo {author}
  {\bibfnamefont {A.~A.}\ \bibnamefont {Aligia}}, \bibinfo {author}
  {\bibfnamefont {C.~A.}\ \bibnamefont {Balseiro}}, \bibinfo {author}
  {\bibfnamefont {G.~K.-L.}\ \bibnamefont {Chan}}, \bibinfo {author}
  {\bibfnamefont {H.~D.}\ \bibnamefont {Abru{\~n}a}}, \ and\ \bibinfo {author}
  {\bibfnamefont {D.~C.}\ \bibnamefont {Ralph}},\ }\href {\doibase
  10.1126/science.1186874} {\bibfield  {journal} {\bibinfo  {journal}
  {Science}\ }\textbf {\bibinfo {volume} {328}},\ \bibinfo {pages} {1370}
  (\bibinfo {year} {2010})}\BibitemShut {NoStop}%
\bibitem [{\citenamefont {Mahan}(1990)}]{Mahan}%
  \BibitemOpen
  \bibfield  {author} {\bibinfo {author} {\bibfnamefont {G.~D.}\ \bibnamefont
  {Mahan}},\ }\href@noop {} {\bibfield  {journal} {\bibinfo  {journal}
  {Many-Particle Physics - Springer}\ ,\ \bibinfo {pages} {227}} (\bibinfo
  {year} {1990})}\BibitemShut {NoStop}%
\bibitem [{\citenamefont {Ziman}(1999)}]{Ziman}%
  \BibitemOpen
  \bibfield  {author} {\bibinfo {author} {\bibfnamefont {J.~M.}\ \bibnamefont
  {Ziman}},\ }\href@noop {} {\bibfield  {journal} {\bibinfo  {journal}
  {Principles of the Theory of Solids - Cambridge University Press}\ ,\
  \bibinfo {pages} {229}} (\bibinfo {year} {1999})}\BibitemShut {NoStop}%
\bibitem [{\citenamefont {Dong}\ and\ \citenamefont {Lei}(2002)}]{Dong_02}%
  \BibitemOpen
  \bibfield  {author} {\bibinfo {author} {\bibfnamefont {B.}~\bibnamefont
  {Dong}}\ and\ \bibinfo {author} {\bibfnamefont {X.~L.}\ \bibnamefont {Lei}},\
  }\href {\doibase 10.1088/0953-8984/14/45/316} {\bibfield  {journal} {\bibinfo
   {journal} {Journal of Physics: Condensed Matter}\ }\textbf {\bibinfo
  {volume} {14}},\ \bibinfo {pages} {11747} (\bibinfo {year}
  {2002})}\BibitemShut {NoStop}%
\bibitem [{\citenamefont {Seridonio}\ \emph
  {et~al.}(2009{\natexlab{b}})\citenamefont {Seridonio}, \citenamefont
  {Yoshida},\ and\ \citenamefont {Oliveira}}]{Seridonio_2009}%
  \BibitemOpen
  \bibfield  {author} {\bibinfo {author} {\bibfnamefont {A.~C.}\ \bibnamefont
  {Seridonio}}, \bibinfo {author} {\bibfnamefont {M.}~\bibnamefont {Yoshida}},
  \ and\ \bibinfo {author} {\bibfnamefont {L.~N.}\ \bibnamefont {Oliveira}},\
  }\href {\doibase 10.1209/0295-5075/86/67006} {\bibfield  {journal} {\bibinfo
  {journal} {{EPL} (Europhysics Letters)}\ }\textbf {\bibinfo {volume} {86}},\
  \bibinfo {pages} {67006} (\bibinfo {year} {2009}{\natexlab{b}})}\BibitemShut
  {NoStop}%
\bibitem [{\citenamefont {Oliveira}\ \emph {et~al.}(2010)\citenamefont
  {Oliveira}, \citenamefont {Yoshida},\ and\ \citenamefont
  {Seridonio}}]{Seridonio_2010}%
  \BibitemOpen
  \bibfield  {author} {\bibinfo {author} {\bibfnamefont {L.~N.}\ \bibnamefont
  {Oliveira}}, \bibinfo {author} {\bibfnamefont {M.}~\bibnamefont {Yoshida}}, \
  and\ \bibinfo {author} {\bibfnamefont {A.~C.}\ \bibnamefont {Seridonio}},\
  }\href {\doibase 10.1088/1742-6596/200/5/052020} {\bibfield  {journal}
  {\bibinfo  {journal} {Journal of Physics: Conference Series}\ }\textbf
  {\bibinfo {volume} {200}},\ \bibinfo {pages} {052020} (\bibinfo {year}
  {2010})}\BibitemShut {NoStop}%
\bibitem [{\citenamefont {Goldhaber-Gordon}\ \emph
  {et~al.}(1998{\natexlab{b}})\citenamefont {Goldhaber-Gordon}, \citenamefont
  {G\"ores}, \citenamefont {Kastner}, \citenamefont {Shtrikman}, \citenamefont
  {Mahalu},\ and\ \citenamefont {Meirav}}]{Goldhaber1998PRL}%
  \BibitemOpen
  \bibfield  {author} {\bibinfo {author} {\bibfnamefont {D.}~\bibnamefont
  {Goldhaber-Gordon}}, \bibinfo {author} {\bibfnamefont {J.}~\bibnamefont
  {G\"ores}}, \bibinfo {author} {\bibfnamefont {M.~A.}\ \bibnamefont
  {Kastner}}, \bibinfo {author} {\bibfnamefont {H.}~\bibnamefont {Shtrikman}},
  \bibinfo {author} {\bibfnamefont {D.}~\bibnamefont {Mahalu}}, \ and\ \bibinfo
  {author} {\bibfnamefont {U.}~\bibnamefont {Meirav}},\ }\href {\doibase
  10.1103/PhysRevLett.81.5225} {\bibfield  {journal} {\bibinfo  {journal}
  {Phys. Rev. Lett.}\ }\textbf {\bibinfo {volume} {81}},\ \bibinfo {pages}
  {5225} (\bibinfo {year} {1998}{\natexlab{b}})}\BibitemShut {NoStop}%
\bibitem [{\citenamefont {Scheibner}\ \emph {et~al.}(2005)\citenamefont
  {Scheibner}, \citenamefont {Buhmann}, \citenamefont {Reuter}, \citenamefont
  {Kiselev},\ and\ \citenamefont {Molenkamp}}]{Scheibner2005}%
  \BibitemOpen
  \bibfield  {author} {\bibinfo {author} {\bibfnamefont {R.}~\bibnamefont
  {Scheibner}}, \bibinfo {author} {\bibfnamefont {H.}~\bibnamefont {Buhmann}},
  \bibinfo {author} {\bibfnamefont {D.}~\bibnamefont {Reuter}}, \bibinfo
  {author} {\bibfnamefont {M.~N.}\ \bibnamefont {Kiselev}}, \ and\ \bibinfo
  {author} {\bibfnamefont {L.~W.}\ \bibnamefont {Molenkamp}},\ }\href {\doibase
  10.1103/PhysRevLett.95.176602} {\bibfield  {journal} {\bibinfo  {journal}
  {Phys. Rev. Lett.}\ }\textbf {\bibinfo {volume} {95}},\ \bibinfo {pages}
  {176602} (\bibinfo {year} {2005})}\BibitemShut {NoStop}%
\bibitem [{\citenamefont {Svensson}\ \emph {et~al.}(2013)\citenamefont
  {Svensson}, \citenamefont {Hoffmann}, \citenamefont {Nakpathomkun},
  \citenamefont {Wu}, \citenamefont {Xu}, \citenamefont {Nilsson},
  \citenamefont {S{\'{a}}nchez}, \citenamefont {Kashcheyevs},\ and\
  \citenamefont {Linke}}]{Svensson2013}%
  \BibitemOpen
  \bibfield  {author} {\bibinfo {author} {\bibfnamefont {S.~F.}\ \bibnamefont
  {Svensson}}, \bibinfo {author} {\bibfnamefont {E.~A.}\ \bibnamefont
  {Hoffmann}}, \bibinfo {author} {\bibfnamefont {N.}~\bibnamefont
  {Nakpathomkun}}, \bibinfo {author} {\bibfnamefont {P.~M.}\ \bibnamefont
  {Wu}}, \bibinfo {author} {\bibfnamefont {H.~Q.}\ \bibnamefont {Xu}}, \bibinfo
  {author} {\bibfnamefont {H.~A.}\ \bibnamefont {Nilsson}}, \bibinfo {author}
  {\bibfnamefont {D.}~\bibnamefont {S{\'{a}}nchez}}, \bibinfo {author}
  {\bibfnamefont {V.}~\bibnamefont {Kashcheyevs}}, \ and\ \bibinfo {author}
  {\bibfnamefont {H.}~\bibnamefont {Linke}},\ }\href@noop {} {\bibfield
  {journal} {\bibinfo  {journal} {New Journal of Physics}\ }\textbf {\bibinfo
  {volume} {15}},\ \bibinfo {pages} {105011} (\bibinfo {year}
  {2013})}\BibitemShut {NoStop}%
\bibitem [{\citenamefont {Merker}\ \emph {et~al.}(2013)\citenamefont {Merker},
  \citenamefont {Kirchner}, \citenamefont {Mu\~noz},\ and\ \citenamefont
  {Costi}}]{Munoz}%
  \BibitemOpen
  \bibfield  {author} {\bibinfo {author} {\bibfnamefont {L.}~\bibnamefont
  {Merker}}, \bibinfo {author} {\bibfnamefont {S.}~\bibnamefont {Kirchner}},
  \bibinfo {author} {\bibfnamefont {E.}~\bibnamefont {Mu\~noz}}, \ and\
  \bibinfo {author} {\bibfnamefont {T.~A.}\ \bibnamefont {Costi}},\ }\href
  {\doibase 10.1103/PhysRevB.87.165132} {\bibfield  {journal} {\bibinfo
  {journal} {Phys. Rev. B}\ }\textbf {\bibinfo {volume} {87}},\ \bibinfo
  {pages} {165132} (\bibinfo {year} {2013})}\BibitemShut {NoStop}%
\bibitem [{\citenamefont {Sato}\ \emph {et~al.}(2005)\citenamefont {Sato},
  \citenamefont {Aikawa}, \citenamefont {Kobayashi}, \citenamefont
  {Katsumoto},\ and\ \citenamefont {Iye}}]{Sato2005}%
  \BibitemOpen
  \bibfield  {author} {\bibinfo {author} {\bibfnamefont {M.}~\bibnamefont
  {Sato}}, \bibinfo {author} {\bibfnamefont {H.}~\bibnamefont {Aikawa}},
  \bibinfo {author} {\bibfnamefont {K.}~\bibnamefont {Kobayashi}}, \bibinfo
  {author} {\bibfnamefont {S.}~\bibnamefont {Katsumoto}}, \ and\ \bibinfo
  {author} {\bibfnamefont {Y.}~\bibnamefont {Iye}},\ }\href {\doibase
  10.1103/PhysRevLett.95.066801} {\bibfield  {journal} {\bibinfo  {journal}
  {Phys. Rev. Lett.}\ }\textbf {\bibinfo {volume} {95}},\ \bibinfo {pages}
  {066801} (\bibinfo {year} {2005})}\BibitemShut {NoStop}%
\end{thebibliography}%

\end{document}